\documentclass[12pt]{article}

\def\thebibliography#1{\section*{References}\list
 {}{\setlength\labelwidth{1.4em}\leftmargin\labelwidth
 \setlength\parsep{0pt}\setlength\itemsep{0pt}
 \setlength{\itemindent}{-\leftmargin}
 \usecounter{enumi}}}

\begin{document}
\begin{center}
{\bf Giant Radio Sources in View of the Dynamical Evolution of FRII-type
Population. I. The Observational Data, and Basic Physical Parameters of
Sources Derived from the Analytical Model}
\end{center}

\begin{center}
by

{J. Machalski, K.T. Chy\.{z}y \& M. Jamrozy}
\end{center}

\begin{center}
Astronomical Observatory, Jagellonian University, ul. Orla 171,\\ 30--244 Cracow, Poland \\
e--mail: machalsk@oa.uj.edu.pl jamrozy@oa.uj.edu.pl
\end{center}

\begin{abstract}

The time evolution of {\sl giant} lobe-dominated radio galaxies
(with projected linear size $D>1$ Mpc if
$H_{0}$=50 km\,s$^{-1}$Mpc$^{-1}$ and $q_{0}$=0.5) is analysed on the basis
of dynamical evolution of the entire FRII-type population. Two
basic physical parameters, namely the jet power $Q_{0}$ and central density of
the galaxy nucleus $\rho_{0}$ are derived for a sample of {\sl giants} with
synchrotron ages reliably determined, and compared with the
relevant parameters in a comparison sample of normal-size sources consisting of
3C, B2, and other sources. Having the apparent radio luminosity $P$ and linear
size $D$ of each sample source, $Q_{0}$ and $\rho_{0}$ are obtained by fitting the
dynamical model of Kaiser et al. (1997). 
We find that:
(i) there is not a unique factor governing the source size; they are old sources
with temperate jet power ($Q_{0}$) evolved in a relatively low-density 
environment ($\rho_{0}$). The size is dependent, in order of decreasing partial
correlation coefficients, on age; then on $Q_{0}$; next on $\rho_{0}$. (ii) A
self-similar expansion of the sources' cocoon seems to be feasible if the power
supplied by the jets is a few orders of magnitude above the minimum-energy value.
In other cases the expansion can only initially be self-similar; a departure from
self-similarity for large and old sources is justified by observational data of
{\sl giant} sources. (iii) An apparent increase 
of the lowest internal pressure value observed within the largest sources' cocoon
with redshift is obscured by the intrinsic dependence of their size on age and the
age on redshift, which hinders us from making definite conclusions about a
cosmological evolution of intergalactic medium (IGM) pressure.

{\bf Key words:} {galaxies: active -- galaxies: evolution -- galaxies: kinematics 
and dynamics}

\end{abstract}

\section{Introduction}

Extragalactic radio sources, powered by twin jets resulting from nuclear
energy processes in the Active Galactic Nucleus (AGN), exhibit a very large
range of their linear size. The sizes of these powerful sources range from
less than $10^{2}$ pc (GPS: Giga Hertz-peaked spectrum), to $10^{2}-10^{4}$
pc (CSS: compact steep spectrum), $10^{4}-10^{6}$ pc (normal-size sources), up 
to greater than $10^{6}$ pc $\equiv 1$ Mpc (`giant' radio sources). One of the 
key problems of the evolution of extragalactic sources is whether and how 
different
size sources are related. Is there a single evolutionary scheme governing the
size evolution of radio sources, or do small and large sources evolve in a
different way?

From many years `giant'\footnote{hereafter we use {\sl giant} or {\sl giants}
instead of giant radio source(s)} radio sources have been of special interest for
several reasons. Their very large angular sizes give excellent opportunity for
the study of source physics. They are also very useful to study the density and
evolution of the intergalactic and intracluster environment (cf. Subrahmanyan 
and Saripalli 1993;
Mack et al. 1998), as well as to verify the unification scheme for powerful
radio sources (Barthel 1989; Urry and Padovani 1995). Finally, they can be used
to constrain dynamical models of the source lifetime evolution (e.g. Kaiser and
Alexander 1999). The general questions are: do the largest radio sources reach
their extremal {\sl giant} sizes due to (i) exceptional physical conditions
in the intergalactic medium, (ii) extraordinary intrinsic properties of the AGN,
or simply (iii) because they are extremely old?

To answer these questions, in a number of papers attempts were made
to recognize properties other than  size which may differentiate {\sl
giants} from normal-size sources. The {\sl giant}-source morphologies, energy
density in the lobes and interaction with the intergalactic medium were studied
by Subrahmanyan et al. (1996), who suggested that {\sl giant} radio galaxies may
be located in the lowest density regions, and `may attained their large size as
a result of restarting of their central engines in multiple phases of activity
along roughly similar directions'. Also Mack et al. (1998), after a study of 5
nearby {\sl giant} radio galaxies, argued that those sources are so huge because
of their low-density environment and not because of their old ages. A similar
conclusion was drawn by Cotter (1998) after his study of a sample of
high-redshift 7C giants. He found those sources to be not old and of similar
kinetic powers of the jet as normal-size 3C sources. But the densities of their
surrounding medium was found to be much lower than those around most of 3C sources
(Rawlings and Saunders 1991). Ishwara-Chandra and Saikia (1999) compiled a sample
of more than 50 known {\sl giant} sources (many of them being of FRI-type 
morphology)
and compared some of their properties with those of a complete sample of 3CR
sources `to investigate the evolution of giant sources, and test their
consistency with the unified scheme for radio galaxies and quasars'. They
concluded that the location of {\sl giants} on the power--linear size ($P-D$)
diagram may suggest that the largest sources have evolved from the smaller
sources. Finally, in the recent extensive study of 26 {\sl giant} galaxies by
Schoenmakers et al. (2000), the authors argued that those galaxies `are both
old sources, in term of their spectral age, and are situated in a relatively
low-density environment, but also that neither of these two properties are
extreme. Therefore, their large size probably results from a combination of
these properties'.

From the above results, it is clear that the phenomenon of the {\sl giant} radio
sources is still open to further research. Therefore, in this paper we analyse
whether observed properties of {\sl giant} sources can be explained by a model
of the dynamical evolution of classical double radio sources in cosmic time, and
what factor (if there is a one) is primarily responsible for the {\sl giant} size.
Obviously such an analysis, based on the giant radio sources only, i.e. the
sources with a strongly
limited range of linear sizes, would not be reliable and even possible. Therefore,
to solve the above problems we analyse the sample of radio sources comprising both
the `giant'-size and `normal'-size sources, most of the latter with sizes from
$\sim$100 kpc to 1 Mpc.

Most published analytical models of the dynamical evolution of extended powerful
radio sources are based on the hydrodynamical self-similar expansion of the
sources' cocoon caused by the interaction of light and supersonic jets with
the ambient medium. Carvalho and O'Dea (2002) classify those models into three
mutually exclusive types encoded as I, II, and III, and give an excellent
description of their properties. However, we realize that only few of them deal
with the source's energetics and the radio luminosity evolution with time which
is crucial for e.g. the analysis of evolutionary tracks of sources on the
luminosity--size ($P$--$D$) plane, i.e. the most sensitive characteristics of the
dynamical models.

Two type III models, published by Kaiser, Dennett-Thorpe and Alexander (1997)
[hereafter referred to as KDA] and Blundell, Rawlings and Willott (1999) [hereafter
BRW], are more sophisticated than models of type I and II. We
apply the KDA model for its simplicity in comparison with the BRW one, and in
spite of some objections about an application of the self-similar models to large
and old sources in which an internal pressure cannot be always above the external
medium pressure (e.g. Hardcastle and Worrall 2000).

The present analysis is confined to classical double radio sources with FRII-type
(Fanaroff and Riley 1974) morphology. Synchrotron ages in a sample comprising
both giant and normal-size sources are used to verify the dynamical time evolution
of such sources predicted by the above analytical model.
Basic physical parameters, i.e. the jet power $Q_{0}$, the central density of the
galaxy nucleus $\rho_{0}$, the energy
density and pressure in the lobes/cocoon ($u_{c}$ and $p_{c}$), and the total
energy of the source $E_{\rm tot}$ are derived with the KDA model for
each member of
the sample to fit its estimated age, redshift, monochromatic radio luminosity,
projected size, and axial ratio. Besides, the fitted values of some of these
parameters are directly compared with the relevant values calculated straight
from the data, e.g. the equipartition magnetic field strength, $B_{\rm eq}$,
equipartition energy density in the source, $u_{\rm eq}$, and its total energy
$U_{\rm eq}$. These `observational' values and their errors are homogeneously
calculated using the method outlined by Miley (1980). Next, the physical
parameters derived for real radio sources are used to specify the conditions or
circumstances under which the FRII-type radio sources can reach the largest
observed linear sizes, to verify the applicability of the self-similar analytical
models for the largest sources, and to search for an evidence after a cosmological
evolution of the ambient pressure in the intergalactic medium (IGM).
 
The observational data used and the physical parameters of the sample sources
found directly from the data are given in Section~2. The application of the
dynamical models is described in Section~3, while in Section~4 results of the
modelling are presented in the form of statistically significant correlations
between physical parameters of the sample sources derived from the fitting
procedure and their apparent observational parameters. The results obtained
are discussed in Section~5 and the conclusions are given in Section~6.

\section{Observational Data}
\subsection[section]{Selection Criteria}

Similarly to the approach of Ishwara-Chandra and Saikia, we have compiled a subsample 
of 18 {\sl giant} sources and a comparison subsample of 54 normal-size sources.
The selection criteria were as follows: (i) the sources have the FRII-type
morphology, (ii) The existing radio maps enable a suitable
determination of their lateral extend, i.e. transversal to the source's axis, and
(iii) their spectral age or the expansion velocity, determined with the same model
of the energy losses, are available from the literature. The spectral ageing data
calculated with the JP model (Jaffe and Perola 1973; cf. Sect.~2.4) for the
{\sl giants} are taken from the papers of Saripalli et al. (1994), Parma et al.
(1996), Mack et al. (1998), Schoenmakers et al. (1998, 2000), Ishwara-Chandra 
and Saikia (1999), Lara et al. (2000), and Machalski and Jamrozy (2000). For the
aim of this paper, i.e. for an observational verification of the dynamical time
evolution of classical radio sources, especially the growth of their linear size
with time predicted by the analytical models,
the comparison subsample has been chosen to comprise high-luminosity (high- and
low-redshift), as well as low-luminosity normal-size sources. The high-redshift
(with $z\geq$0.5) and low-redshift ($z<$0.5) sets consist of 3C sources taken 
from the papers of Alexander and Leahy (1987), Leahy et al. (1989), Liu et al. 
(1992), and Guerra et al. (2000). All of them have
$P_{178}\geq 10^{25}$\,W\,Hz$^{-1}$sr$^{-1}$
(other selection criteria are summarized in Liu et al.). The low-luminosity
set comprises FRII-type sources with $P_{1.4}< 10^{24.4}$
W\,Hz$^{-1}$sr$^{-1}$ (corresponding to $P_{178}<10^{25}$\,W\,Hz$^{-1}$sr$^{-1}$
assuming a mean spectral index of 0.7 between 178 and 1400 MHz). A limited
number of such sources with spectral ages determined have been available from
the papers of Klein et al. (1995) and Parma et al. (1999).

\subsection[section]{Observational Parameters}

For all individual sample sources the following observational parameters are
determined: the redshift $z$, the 1.4 GHz luminosity in W\,Hz$^{-1}$sr$^{-1}$,
the projected linear size $D$ in kpc, the cocoon's axial ratio $AR=D/b$ and
volume $V_{\rm o}$ in kpc$^{3}$. The volume of the source, $V_{\rm o}$, is
calculated assuming a cylindrical geometry with the length $D$, and the base
diameter $b$ taken as the average of the full deconvolved widths of the two lobes.
The latter are measured between 3$\sigma$ contours on a radio contour map 
half-way between the core and the hot spots or distinct extremities of the
source. All these data for the {\sl giant}-size and normal-size sources in our
sample are given in columns 3--7 of Table~1. The columns 
8--9 of Table~1 give the reference papers to the radio map used to determine both
$D$ and $AR$ for each sample source and its spectral age, respectively.

\subsection[section]{Physical Parameters Derived Directly from the Data}

As mentioned in Section~1, the equipartition magnetic field $B_{\rm eq}$,
the energy density $u_{\rm eq}$, and the total emitted energy $U_{\rm eq}$ with
their errors are calculated with the method outlined by Miley (1980).  However,
the Miley's assumption of pure power-law radio spectrum has been abandoned; the
cocoon's radio spectrum has been determined by the least-square method fit of
the simple analytic functions: $y=a+bx+c\exp(\pm x)$ or $y=a+bx+cx^{2}$ (where 
$x=\log\nu$[GHz], $y=\log S(\nu)$) to the available flux densities $S(\nu)$
weighted by their given error. The total luminosity of the cocoon has been
then integrated between 10 MHz and 100 GHz using the above fitted spectrum
with $H_0=50$ km\,s$^{-1}$Mpc$^{-1}$ and $q_0$=0.5\footnote{An application of
the most recent cosmological constants will change numerical values of dimensions,
power, ambient density, etc., but not relations between observational and
physical parameters of the sources. The applied constants provide an easier
comparison of the derived physical parameters of sources with those found in a
large number of previously published papers}.
The values of $u_{\rm eq}$, $B_{\rm eq}$, and
$U_{\rm eq}$=$u_{\rm eq}V_{\rm o}$ with their estimated error, calculated 
for each od the sample sources with the assumption of a filling
factor of unity and equipartition of energy between electrons and protons, are
given in columns 3--5 of Table~2, respectively. Table~2 contains the adopted age 
and physical parameters of the sample sources derived either from the observational 
data (columns 3, 4, and 5) and from the analytical KDA model (columns 6, 7, 8, 
and 9).

\subsection[section]{Spectral Age}

Determination of the age of the radio sources is crucial to constrain any
dynamical model of their time evolution. An apparent age of sources can be
estimated from the ratio of their total emitted energy, $U_{\rm eq}$, determined under the
`minimum energy' condition and the observed power

\[t_{\rm max}\approx U_{\rm eq}/(dU/dt)\,\,\,\,\,\,{\rm where}\,\,\,\,\,\,
dU/dt\equiv \int_{\nu_{1}}^{\nu_{2}} S(\nu)d\nu,\]

\noindent
and $S(\nu)$ is the observed flux density at different frequencies.
The resultant age of source, being rather its upper limit, is usually greater
than the synchrotron age of relativistic particles commonly determined from the
standard spectral-ageing analysis (e.g. Alexander and Leahy 1987). However, the
time-dependence of various energy losses suffered by the particles causes
different parts of the lobes or cocoon to have different ages. Besides, the
radiation losses (and thus the synchrotron age) depend on the history of
particle injection, the distribution of the pitch angle, etc. described by the
different synchrotron models: Kardashev--Pacholczyk (KP); Jaffe and Perola (JP);
or continuous injection (CI), cf. Carilli et al.(1991) for a detailed
description.  Therefore, these models of different energy losses give
different spectral age estimates, and a comparison of ages of the sample sources
ought to be made within the same synchrotron model. Moreover, the synchrotron ages
($t_{\rm syn}$) usually differ from the dynamical age ($t_{\rm dyn}$) estimated
from the ram-pressure arguments (cf. Begelman and Cioffi 1989; Lara et al. 2000;
Schoenmakers et al. 2000). 

An attempt to minimize the discrepancies between spectral and
dynamical ages has been undertaken by Kaiser (2000). His 3-dimensional model of
the synchrotron emissivity of the cocoon traces the individual evolution of
parts of the cocoon and provides, according to the author, a more accurate
estimate for the age of a source. Its application to the lobes of Cygnus~A gave
a very good fit to their observed surface brightness distribution. However,
since an application of the above model is confined to the sources for which their
lobes are reasonably resolved in the direction perpendicular to the jet
axis -- we cannot use it for our statistical approach to the dynamical evolution
of {\sl giant} radio sources and have to rely on results of the standard
ageing analysis.

Basing on a commonly accepted assumption about the proportionality of the spectral
and dynamical ages, hereafter we assume $t_{\rm dyn}=2t_{\rm syn}$ (e.g. Lara et al.
2000). This age (marked by $t$[Myr]) is given in column 2 of Table~2.

\section{Application of the KDA Model}

\subsection[section]{Source Dynamics}

The overall dynamics of a FRII-type source (precisely: its cocoon) described
in the KDA model is based on the earlier self-similar model of Kaiser and Alexander
(1997) [hereafter referred to as KA]. It is assumed that the radio structure is
formed by two jets emanating from the AGN into a
surrounding medium in two opposite directions, then terminating in strong
shocks, and finally inflating the cocoon. A density distribution of the
unperturbed external gas is approximated by a power-law relation
$\rho_{\rm d}=\rho_{0}(d/a_{0})^{-\beta}$, where $d$ is the radial distance from
the core of a source, $\rho_{0}$ is the density at the core radius $a_{0}$, and
$\beta$ is the exponent in this distribution [the simplified King (1972) model].

Half of the cocoon is approximated by a cylinder of length $L_{\rm j}=D_{\rm s}/2$
and axial ratio $R_{\rm T}=AR/2$, where $D_{\rm s}$ is its total unprojected
linear size. The cocoon expands along the jet axis driven by the hot
spot pressure $p_{\rm h}$ and in the perpendicular direction by the cocoon
pressure $p_{\rm c}$. In the model the rate at which energy is transported
along each jet ($Q_{0}$) is constant during the source lifetime.
The model predicts self-similar expansion of the cocoon and gives analytical
formulae for the time evolution of various geometrical and physical parameters,
e.g. the length of the jet [cf. equations (4) and (5) in KA]:

\begin{equation}
L_{\rm j}=D_{\rm s}/2=c_{1}\left(\frac{Q_{0}}{\rho_{0}a_{0}^{\beta}}\right)^{1/(5-\beta)}
t^{3/(5-\beta)};
\end{equation}

\noindent
and the cocoon pressure (cf. equation 34 in KA):

\begin{equation}
p_{\rm c}=\frac{18c_{1}^{2(5-\beta)/3}}{(\Gamma_{\rm x}+1)(5-\beta)^{2}\cal{P_{\rm hc}}}
(\rho_{0}a_{0}^{\beta}Q_{0}^{2})^{1/3}L_{\rm j}^{-(4+\beta)/3},
\end{equation}

\noindent
where $c_{1}$ is a dimensionless constant [equation (25) in KA], and $\Gamma_{\rm x}$
is the adiabatic index  of the unshocked medium surrounding the cocoon,
and ${\cal P}_{\rm hc}\equiv p_{\rm h}/p_{\rm c}$ is the pressure ratio.

However, the pressure ratio ${\cal P}_{\rm hc}=4R_{\rm T}^{2}$, implied in the
original KDA paper, has
later been found to seriously overestimate the value of ${\cal P}_{\rm hc}$
obtained in hydrodynamical simulations by Kaiser and Alexander (1999). Therefore,
in our modelling procedure we use the empirical formula taken from Kaiser (2000):

\begin{equation}
{\cal P}_{\rm hc}=(2.14-0.52\beta)R_{\rm T}^{2.04-0.25\beta}.
\end{equation}

\subsection[section]{Source Energetics and Radio Power}

In our application of the KDA model we neglect thermal particles, hence the 
overall source dynamics is governed by the pressure in the cocoon in the form
$p_{\rm c}=(\Gamma_{\rm c}-1)(u_{\rm e}+u_{\rm B})$, where $\Gamma_{\rm c}$ is
the adiabatic index of the cocoon, $u_{\rm e}$ and $u_{\rm B}$ are the energy
densities of relativistic particles and the magnetic field, respectively. Both
energy densities are a function of the source lifetime $t$. In particular,

\begin{equation}
u_{\rm B}(t)\propto B^{2}(t)={\rm const}\,t^{-a},
\end{equation}

\noindent
where $a=(\Gamma_{\rm B}/\Gamma_{\rm c})(4+\beta)/(5-\beta)$.

Since the time evolution of the  pressure $p_{\rm c}$ is known from the
self-similar solution, then one can calculate the energy density in the cocoon
at any specific age $t$:

\begin{equation}
u_{\rm c}(t)\equiv u_{\rm e}(t)+u_{\rm B}(t)=p_{\rm c}(t)/(\Gamma_{\rm c}-1),
\end{equation}

\noindent
and the total source energy: $E_{\rm tot}(t)=u_{\rm c}(t)V_{\rm c}(t)$, where
$V_{\rm c}$ is the cocoon volume attained at the age $t$:

\begin{equation}
V_{\rm c}(t)=2\frac{\pi}{4R_{\rm T}^{2}}[L_{\rm j}(t)]^{3}\propto t^{9/(5-\beta)}.
\end{equation}

\noindent
Following KDA and Kaiser (2000) we can write:

\begin{equation}
E_{\rm tot}=u_{\rm c}V_{\rm c}=
\frac{2(5-\beta)}{9[\Gamma_{\rm c}+(\Gamma_{\rm c}-1)({\cal P}_{\rm hc}/4)]-4
-\beta}Q_{0}t.
\end{equation}

\noindent
Thus, the ratio of energy delivered by the twin jets and stored in the cocoon is:

\[\frac{2Q_{0}t}{E_{\rm tot}}=\frac{9\Gamma_{\rm c}-4-\beta}{5-\beta} +
\frac{9(\Gamma_{\rm c}-1)}{4(5-\beta)}{\cal P}_{\rm hc},\]

\noindent
i.e. for given $\Gamma_{\rm c}$ and $\beta$ values, 
this ratio is a function of the pressure ratio ${\cal P}_{\rm hc}$ only.
For $\Gamma_{\rm c}=5/3$ and $\beta=3/2$ we have

\begin{equation}
2Q_{0}t/E_{\rm tot}=2.7+0.43 {\cal P}_{\rm hc}.
\end{equation}

The radio power of the cocoon $P_{\nu}$ is calculated in the KDA model by
splitting up the source into small volume elements and allowing them to evolve
separately. The effects of adiabatic expansion, synchrotron losses, and inverse
Compton scattering on the cosmic microwave background radiation are traced for
these volume elements independently. The total radio emission at a fixed
frequency $\nu$ is then obtained by summing  up the contribution from all such
elements, resulting in an integral over time [equation (16) in KDA]. It depends
on the source's age $t$ and redshift $z$, the jet power $Q_{0}$, the cocoon
axial ratio $R_{\rm T}$, the exponent in the expected power-law distribution of
relativistic particles $p$, and on the ratio $r$ of the magnetic field energy
to the energy of relativistic electrons and non-radiating particles (given in
Sect.~3.3). 
The integral is not analytically solvable and,  following the KDA we
calculate it numerically.

\subsection{Fitting Procedure}

On the basis of the above model we aim to predict the specific physical
parameters for all {\sl giants} and normal-size sources in the sample at their
estimated (dynamical) age, i.e. $Q_{0}$, $\rho_{0}$, $u_{\rm c}$, $p_{\rm c}$,
and $E_{\rm tot}$. This differs from the KDA approach, who on the base of
available observational data, evaluated some general trends and made crude
estimates of possible ranges of values attained by the model parameters.

In order to derive the above parameters for our sources, all other free
parameters of the model ($r$, $p$, $\Gamma_{\rm x}$, $\Gamma_{\rm c}$,
$\Gamma_{\rm B}$, $a_{0}$, $\beta$), and the inclination angle of the jet axis
to the observer's line-of-sight $\theta$ have to be approximated.
Following the KDA, we adopt their `Case~3' where both the cocoon and magnetic
field are `cold' ($\Gamma_{\rm c}=\Gamma_{\rm B}=5/3)$ and the adiabatic index
of the jet material and external gas is also $5/3$. For the initial ratio of
the energy densities of the magnetic field and the particles we use
$r\equiv u_{\rm B}/u_{\rm e}=(1+p)/4$, with the exponent of the energy
distribution $p=2.14$.

The core radius $a_{0}$ is one of the most difficult model parameter to be set
up. Even careful 2-D modelling of a distribution of radio emission  for well
known sources with quite regular structures can lead to values of $a_{0}$
discrepant with those predicted by X-ray observations, the only presently
available method to determine the source environment (cf. an extensive discussion
of this problem in Kaiser 2000). In our statistical approach we assume
$a_{0}=10$ kpc for all sources, a conservative value between 2 kpc used by
KDA and 50 kpc found by Wellman et al. (1997). In Section~5.1 we discuss
the consequences of other possible values of this parameter. 

We also use a constant value of $\beta$ for all sample sources taking
$\beta=1.5$ for further calculations. This is compatible with other estimates of
this parameter (e.g. Daly 1995) although much flatter than $\beta=1.9$ adopted in 
the original KDA paper on the basis of Canizares et al.'s (1987) paper who found
that value to be typical for a galaxy at about 100 kpc from its centre. A flatter
density profile should be more adequate for distances of a few hundreds of kpc.

Another free parameter of the model, the orientation of the jet axis with respect
to the observer's line-of-sight $\theta=90^{\circ}$ is assumed for all {\sl giants}
and $\theta=70^{\circ}$ for other sources. This latter value is justified by the 
dominance of FRII-type radio galaxies in our sample. In view of the unified 
scheme for extragalactic radio sources, an average orientation angle 
$\langle\theta_{\rm RG}\rangle\simeq 69^{\circ}$ for radio galaxies only was 
determined by Barthel (1989). The apparent linear size $D$ of a radio source then
yields the model cocoon size:

\begin{equation}
L_{\rm j}=\frac{D}{2\sin\theta}
\end{equation}

Having fixed all these free parameters of the model, we find the jet power 
$Q_{0}$ and the initial density of external medium $\rho_{0}$ for each 
individual sample source by iterative solution of the system of two equation:
(i) equation (1) equated to equation (9) for the jet length $L_{\rm j}$ , and
(ii) the integral for
the luminosity of the cocoon $P_{v}$ [equation (16) in KDA; cf. Section~3.1.2] 
-- requiring the match of the solution to the observed values of $D$ and 
$P_{1.4}$, respectively. The above fitting procedure proved to give always 
 stable and unique solutions. Then, from equations (2) and (5) we calculate 
other model parameters: the cocoon's pressure $p_{\rm c}$ and its energy
density $u_{\rm c}$, and from equation (7) the total cocoon energy 
$E_{\rm tot}$. The resultant values of $Q_{0}$, $\rho_{0}$ and $p_{\rm c}$
for the sample sources are given in columns 6, 7, and 8 of Table~2.

In Table~3 we summarize meanings of the observational and model parameters
characterizing the source's cocoon and used in the present analysis. All
dimensions are given in the SI units except the cocoons' length and volume which
are given in kpc and kpc$^{3}$, respectively. Two quantities present in the text
do not appear in Table~3: the source (cocoon) unprojected linear size $D_{\rm s}$
and unprojected volume $V_{\rm c}$ which imply from the apparent (observed) size
$D$, axial ratio $AR$, and assumed inclination of the sources' axis $\theta$.

\begin{table*}[htb]
\footnotesize
\caption{Data of the sample sources}
\begin{tabular*}{165mm}{@{}lllcrccrl}
\hline
IAU  & Other &$z$& lg$P_{1.4}$ & $D\pm\Delta D$ &AR$\pm\Delta $AR &
lg$V_{\rm o}\pm\Delta$lg$V_{\rm o}$ & Ref. & Spect.\\
name & name  &   & [WHz$^{-1}$sr$^{-1}$] & [kpc] & & [kpc$^{3}$] & map & anal.\\
\hline
GIANTS\\
0109+492   & 3C35     & 0.0670 & 24.53 & 1166$\pm$31 & 3.2$\pm$0.7 & 8.08$\pm$0.17 & 28& 24\\
0136+396   & B2       & 0.2107 & 25.21 & 1555$\pm$35 & 6.0$\pm$1.2 & 7.91$\pm$0.16 & 4 & 6\\
0313+683   & WNB      & 0.0901 & 24.41 & 2005$\pm$34 & 4.2$\pm$0.5 & 8.55$\pm$0.10 & 28& 23\\
0319$-$454 & PKS      & 0.0633 & 24.70 & 2680$\pm$30 & 4.0$\pm$0.8 & 8.98$\pm$0.16 & 22& 22\\
0437$-$244 & MRC      & 0.84   & 26.15 & 1055$\pm$17 & 7.8$\pm$1.5 & 7.18$\pm$0.15 & 5 & 5\\
0813+758   & WNB      & 0.2324 & 25.10 & 2340$\pm$80 & 5.0$\pm$0.5 & 8.60$\pm$0.08 & 25&24\\
0821+695   & 8C       & 0.538  & 25.28 & 2990$\pm$22 & 5.9$\pm$1.0 & 8.77$\pm$0.14 & 7 & 7\\
1003+351   & 3C236    & 0.0988 & 24.76 & 5650$\pm$75 & 9.4$\pm$1.7 & 9.20$\pm$0.14 & 15& 16\\
1025$-$229 & MRC      & 0.309  & 25.28 & 1064$\pm$17 & 5.2$\pm$0.7 & 7.54$\pm$0.11 & 5 & 5\\
1209+745   & 4C74.17  & 0.107  & 24.42 & 1090$\pm$13 & 2.4$\pm$0.5 & 8.25$\pm$0.16 & 2 & 24\\
1232+216   & 3C274.1  & 0.422  & 26.32 & 1024$\pm$15 & 7.4$\pm$1.6 & 7.19$\pm$0.17 & 8 & 1\\
1312+698   & DA340    & 0.106  & 24.76 & 1085$\pm$12 & 4.4$\pm$0.9 & 7.71$\pm$0.16 & 28& 24\\
1343+379   &          & 0.2267 & 24.42 & 3140$\pm$60 & 7.2$\pm$1.1 & 8.67$\pm$0.14 & 28,14& 13\\
1349+647   & 3C292    & 0.71   & 27.02 & 1073$\pm$16 & 6.2$\pm$1.4 & 7.40$\pm$0.18 & 28& 1\\
1358+305   & B2       & 0.206  & 24.86 & 2670$\pm$60 & 3.6$\pm$0.8 & 9.06$\pm$0.17 & 19& 19\\
1543+845   & WNB      & 0.201  & 24.76 & 1950$\pm$25 & 7.6$\pm$1.4 & 8.00$\pm$0.15 & 28& 24\\
1550+202   & 3C326    & 0.0895 & 25.02 & 2510$\pm$55 & 7.0$\pm$1.1 & 8.40$\pm$0.13 & 28& 16\\
2043+749   & 4C74.26  & 0.104  & 24.86 & 1550$\pm$20 & 4.6$\pm$0.8 & 8.14$\pm$0.14 & 28& 24\\
& & &\\
NORMAL\\
0154+286   & 3C55     & 0.720  & 26.79 &  554$\pm$12 & 6.4$\pm$1.5 & 6.54$\pm$0.18 & 9 & 9\\
0229+341   & 3C68.1   & 1.238  & 27.26 &  414$\pm$10 & 4.4$\pm$1.0 & 6.49$\pm$0.18 & 9 & 9\\
0231+313   & 3C68.2   & 1.575  & 27.30 &  190$\pm$4  & 2.8$\pm$0.6 & 5.86$\pm$0.14 & 9 & 9\\
0404+428   & 3C103    & 0.330  & 26.32 &  564$\pm$12 & 6.7$\pm$1.0 & 6.53$\pm$0.12 & 8 & 1\\
0610+260   & 3C154    & 0.5804 & 26.84 &  376$\pm$10 & 2.9$\pm$0.8 & 6.72$\pm$0.21 & 9 & 9\\
0640+233   & 3C165    & 0.296  & 25.94 &  480$\pm$8  & 3.4$\pm$0.8 & 6.90$\pm$0.18 & 8 & 1\\
0642+214   & 3C166    & 0.246  & 26.66 &  187$\pm$15 & 3.1$\pm$0.6 & 5.76$\pm$0.15 & 8 & 1,29\\
0710+118   & 3C175    & 0.768  & 26.85 &  392$\pm$8  & 3.4$\pm$0.9 & 6.64$\pm$0.20 & 9 & 9\\
0806+426   & 3C194    & 1.184  & 27.13 &  122$\pm$3  & 3.1$\pm$0.5 & 5.18$\pm$0.17 & 30& 30\\
0828+324   & B2       & 0.0507 & 24.36 &  396$\pm$14 & 3.2$\pm$0.4 & 6.71$\pm$0.10 & 12& 6,20\\
0908+376   & B2       & 0.1047 & 24.39 &  100$\pm$8  & 2.2$\pm$0.2 & 5.24$\pm$0.08 & 18& 20\\
0958+290   & 3C234    & 0.1848 & 25.84 &  460$\pm$8  & 4.6$\pm$1.0 & 6.59$\pm$0.17 & 8 & 1\\
1008+467   & 3C239    & 1.786  & 27.51 &   94$\pm$3  & 2.6$\pm$0.7 & 5.01$\pm$0.21 & 10& 10\\
1012+488   & GB/GB2   & 0.385  & 26.17 &  694$\pm$13 & 2.2$\pm$0.3 & 7.76$\pm$0.11 & 12& 29\\
1030+585   & 3C244.1  & 0.428  & 26.47 &  352$\pm$7  & 5.4$\pm$1.3 & 6.10$\pm$0.19 & 8 & 1\\
1056+432   & 3C247    & 0.749  & 26.85 &  105$\pm$4  & 3.1$\pm$0.7 & 5.00$\pm$0.18 & 10& 10\\
1100+772   & 3C249.1  & 0.311  & 25.94 &  247$\pm$24 & 2.8$\pm$1.0 & 6.21$\pm$0.26 & 9 & 9\\
1111+408   & 3C254    & 0.734  & 26.85 &  107$\pm$4  & 2.5$\pm$0.3 & 5.22$\pm$0.10 & 10& 10\\
1113+295   & B2       & 0.0489 & 24.21 &   97$\pm$5  & 2.2$\pm$0.3 & 5.20$\pm$0.11 & 18& 20\\
1140+223   & 3C263.1  & 0.824  & 27.31 &   45$\pm$5  & 2.2$\pm$0.4 & 4.20$\pm$0.14 & 10& 10\\
1141+354   & GB/GB2   & 1.781  & 26.68 &   97$\pm$3  & 3.0$\pm$0.4 & 4.93$\pm$0.11 & 11& 29\\
1142+318   & 3C265    & 0.8108 & 27.19 &  644$\pm$16 & 5.4$\pm$1.7 & 6.89$\pm$0.24 & 9 & 1,9\\
1143+500   & 3C266    & 1.275  & 27.15 &   37$\pm$2  & 4.0$\pm$0.4 & 3.42$\pm$0.08 & 10& 10
\end{tabular*}
\end{table*}
\begin{table*}[t]
\footnotesize
\begin{tabular*}{165mm}{@{}lllcrccrl}
\hline
IAU  & Other &$z$& lg$P_{1.4}$ & $D\pm\Delta D$ &AR$\pm\Delta $AR &
lg$V_{\rm o}\pm\Delta$lg$V_{\rm o}$ & Ref. & Spect.\\
name & name  &   & [WHz$^{-1}$sr$^{-1}$] & [kpc] & & [kpc$^{3}$] & map & anal.\\
\hline

1147+130   & 3C267    & 1.144  & 27.17 &  327$\pm$8  & 4.4$\pm$0.7 & 6.18$\pm$0.13 & 9 & 9\\
1157+732   & 3C268.1  & 0.974  & 27.41 &  390$\pm$7  & 4.1$\pm$0.6 & 6.47$\pm$0.12 & 9 & 9\\
1206+439   & 3C268.4  & 1.400  & 27.37 &   87$\pm$4  & 2.8$\pm$0.3 & 4.85$\pm$0.09 & 10& 10\\
1216+507   & GB/GB2   & 0.1995 & 24.93 &  826$\pm$8  & 4.4$\pm$0.6 & 7.39$\pm$0.11 & 12& 29\\
1218+339   & 3C270.1  & 1.519  & 27.48 &  104$\pm$12 & 2.6$\pm$0.5 & 5.15$\pm$0.15 & 10& 10\\
1221+423   & 3C272    & 0.944  & 26.73 &  490$\pm$13 & 3.6$\pm$1.1 & 6.88$\pm$0.23 & 28& 29\\
1241+166   & 3C275.1  & 0.557  & 26.56 &  130$\pm$15 & 2.0$\pm$0.4 & 5.66$\pm$0.16 & 10& 10\\
1254+476   & 3C280    & 0.996  & 27.35 &  110$\pm$13 & 2.4$\pm$0.3 & 5.29$\pm$0.10 & 10& 10\\
1308+277   & 3C284    & 0.2394 & 25.63 &  836$\pm$6  & 6.9$\pm$1.3 & 7.01$\pm$0.15 & 8 & 1\\
1319+428   & 3C285    & 0.0794 & 24.68 &  271$\pm$4  & 2.8$\pm$0.6 & 6.33$\pm$0.17 & 8 & 1\\
1343+500   & 3C289    & 0.967  & 27.02 &   86$\pm$2  & 2.3$\pm$0.2 & 5.00$\pm$0.07 & 10& 10\\
1347+285   & B2       & 0.0724 & 23.60 &   86$\pm$4  & 2.4$\pm$0.3 & 4.94$\pm$0.10 & 18& 20\\
1404+344   & 3C294    & 1.779  & 27.41 &  132$\pm$17 & 3.8$\pm$1.0 & 5.12$\pm$0.20 & 10& 10\\
1420+198   & 3C300    & 0.270  & 26.00 &  516$\pm$6  & 3.0$\pm$1.2 & 7.11$\pm$0.29 & 8 & 1\\
1441+262   & B2       & 0.0621 & 23.51 &  333$\pm$8  & 4.0$\pm$0.9 & 6.28$\pm$0.18 & 21& 20\\
1522+546   & 3C319    & 0.192  & 25.56 &  390$\pm$15 & 3.2$\pm$0.7 & 6.69$\pm$0.17 & 8 & 1\\
1533+557   & 3C322    & 1.681  & 27.49 &  279$\pm$7  & 3.6$\pm$0.7 & 6.15$\pm$0.15 & 9 & 9\\
1547+215   & 3C324    & 1.207  & 27.28 &   88$\pm$3  & 3.6$\pm$0.9 & 4.60$\pm$0.30 & 30& 30\\
1549+628   & 3C325    & 0.860  & 27.09 &  132$\pm$4  & 4.4$\pm$0.9 & 4.97$\pm$0.28 & 30& 30\\
1609+660   & 3C330    & 0.549  & 26.93 &  458$\pm$8  & 6.4$\pm$1.2 & 6.29$\pm$0.15 & 9 & 9\\
1609+312   & B2       & 0.0944 & 23.65 &   56$\pm$4  & 2.4$\pm$0.3 & 4.41$\pm$0.10 & 3 & 20\\
1615+324   & 3C332    & 0.1515 & 25.32 &  306$\pm$7  & 4.8$\pm$1.3 & 6.02$\pm$0.21 & 3 & 20\\
1618+177   & 3C334    & 0.555  & 26.45 &  430$\pm$15 & 2.8$\pm$0.4 & 6.93$\pm$0.12 & 9 & 9\\
1627+444   & 3C337    & 0.635  & 26.70 &  337$\pm$4  & 5.0$\pm$0.9 & 6.08$\pm$0.19 & 30& 30\\
1658+302   & B2       & 0.0351 & 23.39 &  120$\pm$10 & 2.2$\pm$0.2 & 5.47$\pm$0.07 & 21& 20\\
1723+510   & 3C356    & 1.079  & 26.96 &  643$\pm$13 & 7.9$\pm$1.0 & 6.55$\pm$0.10 & 9 & 9\\
1726+318   & 3C357    & 0.1664 & 25.43 &  395$\pm$10 & 3.0$\pm$0.6 & 6.76$\pm$0.16 &21,3&20\\
1957+405   & CygA     & 0.0564 & 27.25 &  185$\pm$3  & 3.8$\pm$0.5 & 5.57$\pm$0.11 & 9 & 9\\
2019+098   & 3C411    & 0.467  & 27.05 &  201$\pm$5  & 2.6$\pm$0.6 & 6.00$\pm$0.18 & 26& 26\\
2104+763   & 3C427.1  & 0.572  & 26.75 &  173$\pm$5  & 2.9$\pm$0.7 & 5.71$\pm$0.19 & 9 & 9\\
2145+151   & 3C437    & 1.48   & 27.43 &  317$\pm$9  & 5.9$\pm$1.0 & 5.86$\pm$0.19 & 30& 30\\
\hline
\end{tabular*}

\vspace{2mm}
\begin{tabular*}{135mm}{@{}rlrl}
{\bf References}\\
(1)& Alexander and Leahy 1987&         (16)& Mack et al. 1998\\
(2)& van Breugel and Willis 1981&      (17)& Myers and Spangler 1985\\
(3)& Fanti et al. 1986&               (18)& Parma et al. 1986\\
(4)& Hine 1979&                       (19)& Parma et al. 1996\\
(5)& Ishwara-Chandra and Saikia 1999&  (20)& Parma et al. 1999\\
(6)& Klein et al. 1995&               (21)& de Ruiter et al. 1986\\
(7)& Lara et al. 2000&                (22)& Saripalli et al. 1994\\
(8)& Leahy and Williams 1984&          (23)& Schoenmakers et al. 1998\\
(9)& Leahy et al. 1989&               (24)& Schoenmakers et al. 2000\\
(10)& Liu et al. 1992&                (25)& Schoenmakers et al. 2001\\
(11)& Machalski and Condon 1983&       (26)& Spangler and Pogge 1984\\
(12)& Machalski and Condon 1985&       (27)& FIRST (Becker et al. 1996)\\
(13)& Machalski and Jamrozy 2000&      (28)& NVSS (Condon et al. 1998)\\
(14)& Machalski et al. 2001&          (29)& this paper\\
(15)& Mack et al. 1997&               (30)& Guerra et al. 2000\\
\end{tabular*}
\end{table*}

\section{Results of the Modelling} 
\subsection[section]{Jet Power $Q_{0}$ and Core Density $\rho_{0}$}

In the KDA model the jet power and 
ambient density are independent in accordance with a physical intuition.
A distribution of these parameters derived for different sets of the sample
sources on the log($Q_{0}$)--log($\rho_{0}$) plane is shown in Figure~1a.
One can realise in Figure~1a that {\sl giants} are not fully separated from
other sources, where (i) among the sources with
a comparable jet power $Q_{0}$, {\sl giant} sources have an average central
density $\rho_{0}$ smaller than a corresponding central density of normal-size
sources, (ii) {\sl giants} have at least ten times more powerful jets than much
smaller low-luminosity sources of a comparable $\rho_{0}$.

Moreover, for a number of sources in the sample the derived values of their 
fundamental parameters $Q_{0}$ and $\rho_{0}$ are very close, while their ages
are significantly different. Thus in view of the model assumptions, they may be
 considered as `the same' source observed at different epochs of its
lifetime. Such bunches of three to five sources (hereafter called `clans') are
indicated in Figure~1a with the large circles. These clans have appeared crucial
in a comparison of the observing data and the model predictions, and in the
analysis of the {\sl giant}-source phenomenon. More detailed analysis of these
clans and their evolution will be given in Paper II of this series.
 
\begin{figure}
\includegraphics{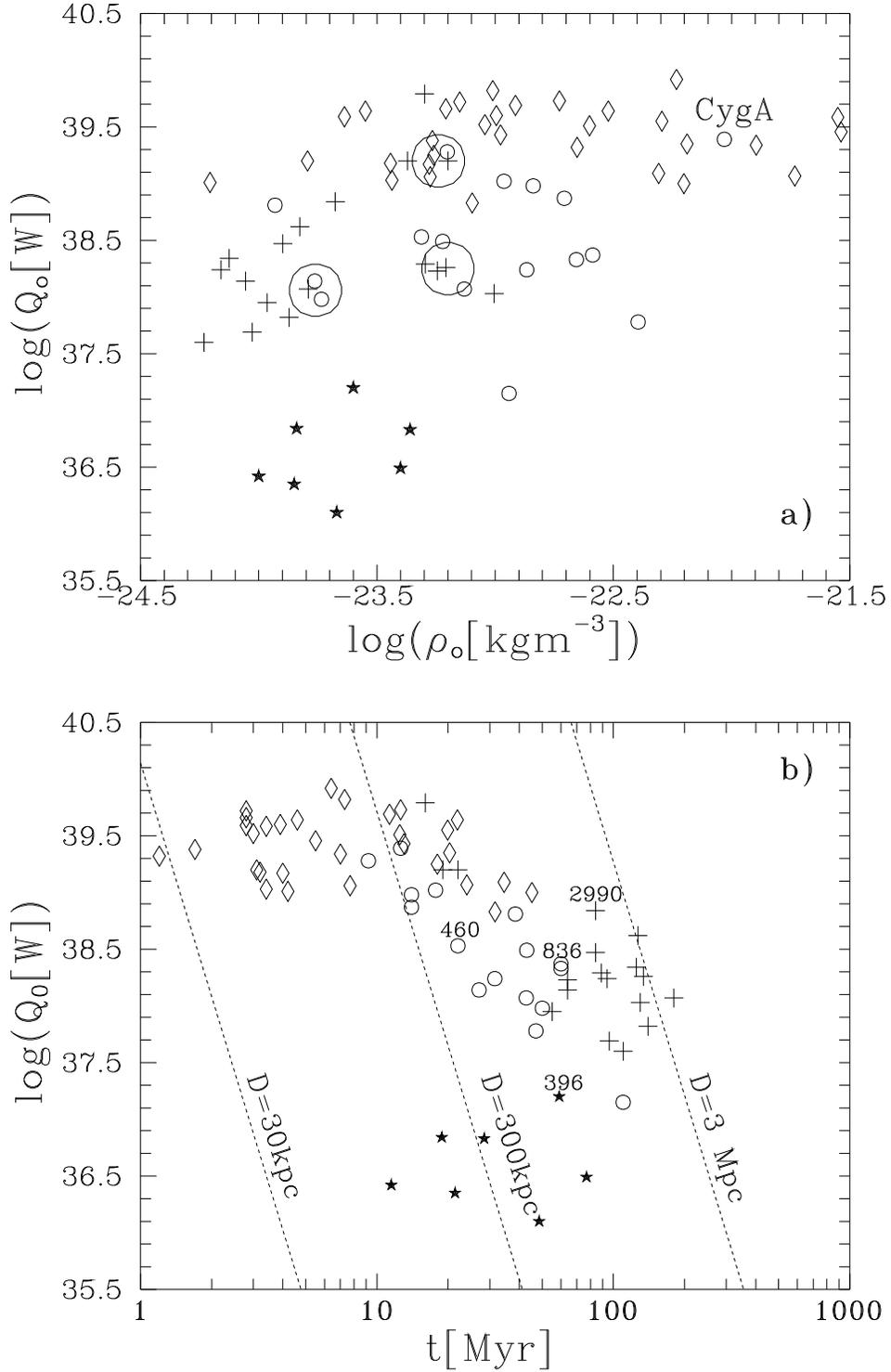}
\vspace{170mm}
\caption{Plots of the jet power $Q_{0}$ {\bf a)} against central density of the
core $\rho_{0}$; {\bf b)} against source age. The {\sl giants} are indicated by
crosses, high-redshift sources --
diamonds, low-redshift sources -- open circles, and low-luminosity sources --
stars. The dotted lines mark
a constant linear size predicted from equation (10). The numbers above some symbols
indicate observed size of the marked source.}
\label{f1}
\end{figure}

The KDA and BWR models predict that the luminosities of mature radio sources decrease
with their age. Therefore, more distant sources fall below the flux-density
limit of a sample sooner than nearer sources, and in any sample the 
high-redshift sources will be younger and more luminous than the low-redshift
ones. A significant anticorrelation between $Q_{0}$ and age $t$, expected as a 
consequence of the above effect (called `youth-redshift degeneracy' in BWR), is
shown in Figure~1b.

\subsection[section]{Cocoon's Energy Density $u_{\rm c}$ and Total Emitted
Energy $E_{\rm tot}$}

In the KDA model, the energetics of the radio source is governed by the jet power
$Q_{0}$, adiabatic index $\Gamma_{\rm c}$, and the pressure ratio ${\cal P}_{\rm hc}$.
Since
$\Gamma_{\rm c}$ is assumed constant for all the sources and $Q_{0}$ is constant
for a given source, the energy of the cocoon $u_{\rm c}$  is determined by the
pressure $p_{\rm c}$ attained by the cocoon at age $t$. As it decreases with time
[cf. Equations (5), (3) and (2)] and the volume increases with time [Equation (6)],
their product, i.e. the model total emitted energy, $E_{\rm tot}$, is
the increasing function of time, and its value is a fraction of energy delivered
by the jet since the source's birth, $Q_{0}t$ [cf. Equation (7)]. A distribution of
$u_{\rm c}$ and $E_{\rm tot}$ parameters on the log($u_{\rm c}$)--log($E_{\rm tot}$)
plane is shown in Figure~2. The time axes, calculated with Equations (2), (5) and
(7) for a constant jet power, are indicated by the dotted lines.

In order to investigate the time evolution of {\sl giant} radio sources
on the basis of dynamical evolution of the entire FRII-type population,
in the next Subsection we examine several correlations between the basic
physical parameters of the sample sources derived from the data and
the above two models. In the result we realize that all statistical
tendencies are similar in both models. Therefore, below we present
these correlations between parameters derived with the preferred
KDA model only.

\begin{figure}
\includegraphics{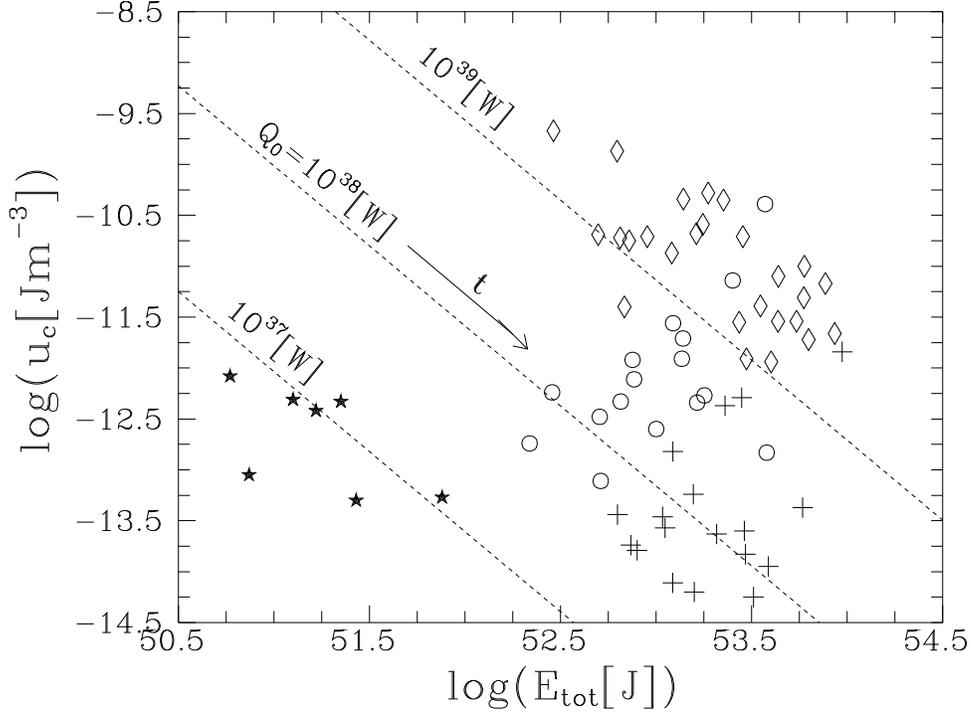}
\vspace{75mm} 
\caption{Plot of the energy density $u_{\rm c}$ against total emitted energy
$E_{\rm tot}$. The symbols indicating sources are the same as in Figure~1. Dotted
lines mark the time axes for different values of $Q_{0}$.}
\label{f2}
\end{figure}

\subsection[section]{Correlations between Observed and Model Parameters}

Below we analyse the relations between principal observational and model
parameters of the sample sources important for their time evolution. Most of
these parameters are interdependent (for example: the linear size of a source is
likely dependent both on its age and the ambient medium density), hence each
parameter of sources in our sample correlates somehow with other parameters.
Therefore, in order to determine which correlation is the strongest, we calculate
the Pearson partial correlation coefficients between selected parameters. For the
reason that most correlations between different parameters seem to be a power law,
all correlations are calculated between logarithms of the given parameters 
(for the sake of simplicity,  the `log' signs are omitted in all 
Tables showing the partial correlations). Hereafter
$r_{XY}$ denotes the correlation coefficient between parameters
$X$ and $Y$, $r_{XY/W}$ is the partial correlation coefficient between these
parameters in the presence of a third one, ($W$), which can correlate with both
$X$ and $Y$, and $P_{XY/W}$ is the probability that the test pair $X$ and $Y$ is
uncorrelated when $W$ is held constant. Similarly, $r_{XY/VW}$, $r_{XY/UVW}$,
$P_{XY/VW}$, and $P_{XY/UVW}$ are the correlation coefficients for the correlations
involving four or five parameters, and the related probabilities, respectively.

A strong correlation between linear size and spectral age of 3C radio sources was
already noted by Alexander and Leahy (1987) and confirmed by Liu et al. (1992).
This correlation in our samples is shown in Figure~3. The  giant sources
do not show any tendency to a faster expansion velocity than that for the
normal-size sources. The same conclusion 
has been made by Schoenmakers et al. (2000). However, two other
aspects are worth emphasizing: (i) There are four high-redshift {\sl giants}
which are much younger than the low-redshift {\sl giants}. Two of them
are quasars. It seems that they might grow so large under some exceptional
conditions. (ii) The $D-t$ relation for the low-luminosity sources (mostly B2) 
follows the same slope of the correlation as that for other sources, but 
low-luminosity sources are definitely much smaller indicating a dependence 
of the size and expansion velocity on the source luminosity.

\begin{figure}[t]
\includegraphics{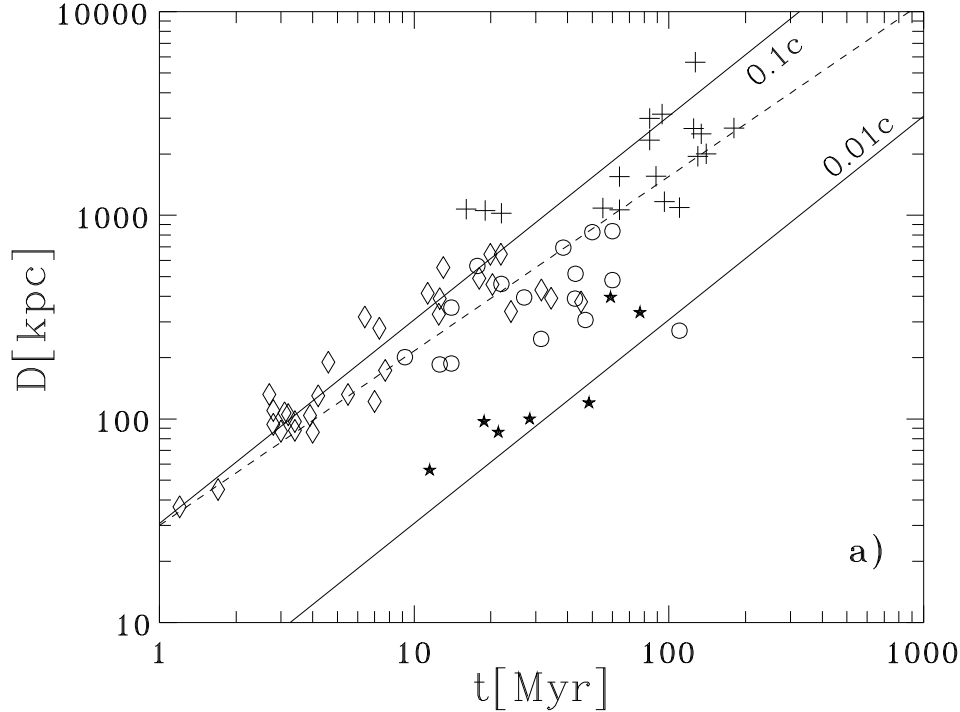}
\vspace{75mm} 
\caption{Plot of the linear size $D$ against source age $t$. The
symbols indicating sources are the same as in Figure~1. The solid lines indicate
the implied expansion velocities in the speed of light $c$ units. The dashed line
shows the model predicted $D(t)\propto L_{\rm j}(t)$ relation resulting from
Equation~(1) for $\beta$=1.5.}
\label{f3}
\end{figure}

The partial correlation coefficients between the size $D$ and  $t$, $Q_{0}$, and
$\rho_{0}$ together with the related probabilities of their chance correlation
are given in Table~4.

\begin{table*}[htb]
\footnotesize
\caption{Age and physical parameters}
\begin{tabular*}{165mm}{@{}lrcccllrr}
\hline
Source & t & lg$u_{\rm eq}$ & $B_{\rm eq}$ & lg$U_{\rm eq}$
&lg$Q_{0}$ & lg$\rho_{0}$ & lg$p_{\rm c}$ &\underline{$2Q_{0}t$}\\
       &[Myr]&[Jm$^{-3}$] & [nT] & [J] & [W] & [kgm$^{-3}$] & [Nm$^{-2}$] & $U_{\rm eq}$\\
\hline
GIANTS\\
0109+492   &  96$\pm$18 & $-13.74\pm 0.12$ & 0.14$\pm$0.02 & 52.81$\pm$0.28 & 37.69 & $-$24.03 & $-$13.86  &4.6$\pm$2.0\\
0136+396   &  89$\pm$17 & $-13.24\pm 0.17$ & 0.25$\pm$0.05 & 53.14$\pm$0.31 & 38.29 & $-$23.30 & $-$13.36  &8.0$\pm$4.2\\
0313+683   & 140$\pm$24 & $-14.11\pm 0.14$ & 0.09$\pm$0.02 & 52.91$\pm$0.23 & 37.82 & $-$23.87 & $-$14.11  &7.2$\pm$2.6\\
0319$-$454 & 180$\pm$40 & $-13.83\pm 0.13$ & 0.13$\pm$0.02 & 53.62$\pm$0.28 & 38.07 & $-$23.79 & $-$14.15  &3.2$\pm$1.3\\
0437$-$244 &  19$\pm$6  & $-12.37\pm 0.15$ & 0.68$\pm$0.12 & 53.28$\pm$0.29 & 39.20 & $-$23.37 & $-$12.46 &10.0$\pm$3.5\\
0813+758   &  84$\pm$4  & $-13.60\pm 0.15$ & 0.17$\pm$0.03 & 53.47$\pm$0.22 & 38.47 & $-$23.90 & $-$13.79  &5.3$\pm$2.5\\
0821+695   &  84$\pm$10 & $-13.37\pm 0.17$ & 0.21$\pm$0.04 & 53.87$\pm$0.30 & 38.84 & $-$23.68 & $-$13.64  &5.0$\pm$2.8\\
1003+351   & 127$\pm$18 & $-14.25\pm 0.14$ & 0.08$\pm$0.01 & 53.42$\pm$0.27 & 38.62 & $-$23.83 & $-$14.34 &12.8$\pm$6.1\\
1025$-$229 &  64$\pm$12 & $-12.82\pm 0.22$ & 0.40$\pm$0.10 & 53.19$\pm$0.31 & 38.23 & $-$23.25 & $-$13.10  &4.4$\pm$2.3\\
1209+745   & 110$\pm$20 & $-13.79\pm 0.13$ & 0.13$\pm$0.02 & 52.93$\pm$0.28 & 37.60 & $-$24.23 & $-$13.99  &5.3$\pm$1.5\\
1232+216   &  22$\pm$3  & $-12.29\pm 0.32$ & 0.74$\pm$0.27 & 53.37$\pm$0.43 & 39.20 & $-$23.20 & $-$12.38  &9.4$\pm$8.0\\
1312+698   &  55$\pm$5  & $-13.44\pm 0.14$ & 0.19$\pm$0.03 & 52.74$\pm$0.29 & 37.95 & $-$23.97 & $-$13.56  &5.7$\pm$3.2\\
1343+379   &  94$\pm$16 & $-14.20\pm 0.33$ & 0.08$\pm$0.03 & 53.07$\pm$0.41 & 38.24 & $-$24.16 & $-$14.25  &8.8$\pm$6.8\\
1349+647   &  16$\pm$4  & $-11.84\pm 0.17$ & 1.25$\pm$0.25 & 54.03$\pm$0.33 & 39.79 & $-$23.30 & $-$12.07  &5.8$\pm$3.0\\
1358+305   & 125$\pm$25 & $-13.95\pm 0.14$ & 0.12$\pm$0.02 & 53.58$\pm$0.30 & 38.34 & $-$24.12 & $-$14.10  &4.6$\pm$2.2\\
1543+845   & 130$\pm$21 & $-13.46\pm 0.23$ & 0.19$\pm$0.05 & 53.01$\pm$0.35 & 38.03 & $-$23.00 & $-$13.61  &8.6$\pm$5.6\\
1550+202   & 134$\pm$27 & $-13.63\pm 0.27$ & 0.16$\pm$0.05 & 53.24$\pm$0.36 & 38.26 & $-$23.21 & $-$13.73  &8.9$\pm$5.7\\
2043+749   &  64$\pm$11 & $-13.57\pm 0.10$ & 0.17$\pm$0.02 & 53.04$\pm$0.23 & 38.14 & $-$24.06 & $-$13.74  &5.1$\pm$1.9\\
& & &\\
NORMAL\\
0154+286   &  13$\pm$2  & $-11.55\pm 0.16$ & 1.74$\pm$0.32 & 53.46$\pm$0.32 & 39.43 & $-$22.98 & $-$11.72  &7.7$\pm$5.0\\  
0229+341   &  11$\pm$2  & $-11.31\pm 0.19$ & 2.31$\pm$0.43 & 53.65$\pm$0.35 & 39.69 & $-$22.91 & $-$11.33  &7.8$\pm$5.4\\
0231+313   & 4.6$\pm$0.3& $-10.71\pm 0.16$ & 4.58$\pm$0.85 & 53.62$\pm$0.29 & 39.64 & $-$23.55 & $-$11.03  &3.2$\pm$1.9\\
0404+428   &  18$\pm$3  & $-11.91\pm 0.17$ & 1.15$\pm$0.23 & 53.08$\pm$0.30 & 39.02 & $-$22.96 & $-$12.01 &10.2$\pm$5.5\\
0610+260   &  45$\pm$4  & $-11.72\pm 0.18$ & 1.43$\pm$0.30 & 53.47$\pm$0.36 & 39.00 & $-$22.20 & $-$11.54 &10.2$\pm$7.6\\
0640+233   &  60$\pm$10 & $-12.34\pm 0.16$ & 0.70$\pm$0.13 & 53.03$\pm$0.32 & 38.33 & $-$22.66 & $-$12.30  &7.9$\pm$4.5\\
0642+214   &  14$\pm$5  & $-11.71\pm 0.22$ & 1.44$\pm$0.37 & 52.58$\pm$0.34 & 38.87 & $-$22.71 & $-$11.23 &17.9$\pm$7.9\\
0710+118   &  35$\pm$5  & $-11.54\pm 0.16$ & 1.76$\pm$0.32 & 53.57$\pm$0.34 & 39.09 & $-$22.31 & $-$11.52  &7.8$\pm$5.0\\
0806+426   &  7.0$\pm$1.2& $-10.46\pm 0.12$ & 6.13$\pm$0.82 & 53.22$\pm$0.33 & 39.27 & $-$22.56 & $-$10.57 &5.0$\pm$2.3\\ 
0828+324   &  59$\pm$9  & $-13.27\pm 0.18$ & 0.24$\pm$0.05 & 51.90$\pm$0.27 & 37.20 & $-$23.60 & $-$13.24  &7.8$\pm$3.6\\
0908+376   &  28$\pm$5  & $-12.33\pm 0.16$ & 0.71$\pm$0.13 & 51.38$\pm$0.23 & 36.83 & $-$23.36 & $-$12.39  &5.5$\pm$2.0\\
0958+290   &  22$\pm$4  & $-12.11\pm 0.15$ & 0.92$\pm$0.16 & 52.94$\pm$0.30 & 38.53 & $-$23.31 & $-$12.32  &5.7$\pm$2.9\\
1008+467   & 2.8$\pm$0.3& $-10.28\pm 0.16$ & 7.48$\pm$1.39 & 53.20$\pm$0.35 & 39.66 & $-$23.21 & $-$10.36  &5.5$\pm$4.0\\
1012+488   &  38$\pm$6  & $-12.83\pm 0.13$ & 0.40$\pm$0.06 & 53.40$\pm$0.23 & 38.81 & $-$23.93 & $-$12.80  &6.7$\pm$2.6\\
1030+585   &  14$\pm$2  & $-11.56\pm 0.16$ & 1.72$\pm$0.32 & 53.01$\pm$0.33 & 38.98 & $-$22.84 & $-$11.62  &8.9$\pm$5.5\\
1056+432   & 3.2$\pm$0.3& $-10.72\pm 0.14$ & 4.55$\pm$0.71 & 52.75$\pm$0.31 & 39.18 & $-$23.44 & $-$10.81  &5.7$\pm$3.5\\
1100+772   &  32$\pm$6  & $-11.92\pm 0.16$ & 1.14$\pm$0.21 & 52.76$\pm$0.38 & 38.24 & $-$22.87 & $-$11.95  &6.5$\pm$4.6\\
1111+408   & 3.1$\pm$0.2& $-10.75\pm 0.15$ & 4.40$\pm$0.76 & 52.94$\pm$0.24 & 39.20 & $-$23.79 & $-$10.97  &3.8$\pm$1.9\\
1113+295   &  19$\pm$3  & $-12.42\pm 0.16$ & 0.64$\pm$0.12 & 51.25$\pm$0.26 & 36.84 & $-$23.84 & $-$12.51  &5.0$\pm$2.2\\
1140+223   & 1.7$\pm$0.6& $-9.87 \pm 0.11$ & 12.0$\pm$1.50 & 52.80$\pm$0.24 & 39.38 & $-$23.27 & $-$10.02  &4.4$\pm$1.0\\
1141+354   & 3.4$\pm$0.8& $-10.69\pm 0.16$ & 4.68$\pm$0.87 & 52.71$\pm$0.26 & 39.03 & $-$23.44 & $-$10.85  &4.8$\pm$2.4\\
1142+318   &  22$\pm$5  & $-11.66\pm 0.16$ & 1.54$\pm$0.29 & 53.69$\pm$0.37 & 39.64 & $-$22.52 & $-$11.57 &13.0$\pm$8.1\\
1143+500   & 1.2$\pm$0.5&  $-9.67\pm 0.16$ & 15.1$\pm$2.80 & 52.22$\pm$0.23 & 39.32 & $-$22.65 &  $-$9.58 &10.0$\pm$4.2\\
\end{tabular*}
\end{table*}

\begin{table*}[t]
\footnotesize
\begin{tabular*}{165mm}{@{}lrcccllrr}
\hline
Source & t & lg$u_{\rm eq}$ & $B_{\rm eq}$ & lg$U_{\rm eq}$
&lg$Q_{0}$ & lg$\rho_{0}$ & lg$p_{\rm c}$ &\underline{$2Q_{0}t$}\\
       &[Myr]&[Jm$^{-3}$] & [nT] & [J] & [W] & [kgm$^{-3}$] & [Nm$^{-2}$] & $U_{\rm eq}$\\
\hline

%\begin{tabular*}{160mm}{@{}lrrlcllrr}
1147+130   &  12$\pm$4  & $-11.10\pm 0.14$ & 2.93$\pm$0.48 & 53.55$\pm$0.26 & 39.51 & $-$22.60 & $-$11.16  &7.7$\pm$2.4\\
1157+732   &  13$\pm$3  & $-11.17\pm 0.16$ & 2.70$\pm$0.49 & 53.77$\pm$0.27 & 39.73 & $-$22.73 & $-$11.20  &7.8$\pm$3.2\\
1206+439   & 3.0$\pm$0.3& $-10.34\pm 0.12$ & 7.00$\pm$1.00 & 52.98$\pm$0.20 & 39.52 & $-$23.04 & $-$10.32  &7.1$\pm$2.6\\
1216+507   &  50$\pm$15 & $-13.11\pm 0.18$ & 0.29$\pm$0.06 & 52.75$\pm$0.28 & 37.98 & $-$23.73 & $-$13.29  &5.8$\pm$1.9\\
1218+339   & 3.9$\pm$1.1& $-10.35\pm 0.17$ & 6.90$\pm$1.35 & 53.26$\pm$0.30 & 39.60 & $-$23.00 & $-$10.41  &5.7$\pm$2.4\\
1221+423   &  18$\pm$8  & $-11.94\pm 0.16$ & 1.12$\pm$0.21 & 53.41$\pm$0.36 & 39.25 & $-$23.26 & $-$11.90  &8.1$\pm$3.1\\
1241+166   & 4.2$\pm$0.4& $-11.40\pm 0.16$ & 2.07$\pm$0.38 & 52.73$\pm$0.30 & 39.01 & $-$24.21 & $-$11.44  &5.5$\pm$3.3\\
1254+476   & 2.8$\pm$0.3& $-10.68\pm 0.16$ & 4.77$\pm$0.88 & 53.08$\pm$0.25 & 39.59 & $-$23.64 & $-$10.69  &6.2$\pm$2.9\\
1308+277   &  60$\pm$9  & $-12.60\pm 0.18$ & 0.52$\pm$0.11 & 52.88$\pm$0.31 & 38.37 & $-$22.59 & $-$12.63 &12.6$\pm$7.2\\
1319+428   & 110$\pm$20 & $-12.77\pm 0.16$ & 0.43$\pm$0.08 & 52.03$\pm$0.30 & 37.15 & $-$22.79 & $-$12.90  &9.1$\pm$4.9\\
1343+500   & 4.0$\pm$0.5& $-10.71\pm 0.16$ & 4.61$\pm$0.85 & 52.76$\pm$0.22 & 39.17 & $-$23.28 & $-$10.67  &7.0$\pm$2.7\\
1347+285   &  21$\pm$4  & $-12.31\pm 0.13$ & 0.73$\pm$0.11 & 51.13$\pm$0.22 & 36.35 & $-$23.85 & $-$12.72  &2.4$\pm$0.8\\
1404+344   & 2.8$\pm$0.2& $-10.59\pm 0.16$ & 5.23$\pm$0.97 & 53.00$\pm$0.34 & 39.72 & $-$23.15 & $-$10.50  &9.8$\pm$6.7\\
1420+198   &  43$\pm$6  & $-12.27\pm 0.14$ & 0.76$\pm$0.12 & 53.31$\pm$0.39 & 38.49 & $-$23.22 & $-$12.47  &4.4$\pm$3.3\\
1441+262   &  77$\pm$13 & $-13.30\pm 0.19$ & 0.23$\pm$0.05 & 51.46$\pm$0.35 & 36.49 & $-$23.40 & $-$13.46  &5.6$\pm$3.5\\
1522+546   &  43$\pm$7  & $-12.33\pm 0.15$ & 0.71$\pm$0.12 & 52.82$\pm$0.30 & 38.07 & $-$23.13 & $-$12.49  &5.1$\pm$2.7\\
1533+557   & 7.3$\pm$0.6& $-11.00\pm 0.16$ & 3.28$\pm$0.60 & 53.62$\pm$0.30 & 39.82 & $-$23.01 & $-$10.99  &7.9$\pm$4.7\\
1547+215   &  3.4$\pm$1.0& $-10.08\pm 0.18$ & 9.44$\pm$1.89 & 53.02$\pm$0.40 & 39.45 & $-$22.62 & $-$10.19 &5.8$\pm$4.1\\ 
1549+628   &  5.5$\pm$1.5& $-10.55\pm 0.15$ & 5.50$\pm$0.90 & 52.91$\pm$0.47 & 39.34 & $-$22.47 & $-$10.51 &9.3$\pm$4.8\\
1609+660   &  20$\pm$3.5& $-11.39\pm 0.18$ & 2.10$\pm$0.43 & 53.37$\pm$0.31 & 39.35 & $-$22.19 & $-$11.36 &12.9$\pm$7.1\\
1609+312   &  12$\pm$3  & $-12.08\pm 0.14$ & 0.95$\pm$0.15 & 50.80$\pm$0.23 & 36.42 & $-$24.00 & $-$12.39  &3.3$\pm$1.0\\
1615+324   &  47$\pm$9  & $-12.24\pm 0.15$ & 0.79$\pm$0.14 & 52.25$\pm$0.34 & 37.78 & $-$22.40 & $-$12.18 &10.8$\pm$6.3\\
1618+177   &  32$\pm$4  & $-11.91\pm 0.15$ & 1.15$\pm$0.20 & 53.49$\pm$0.26 & 38.83 & $-$23.10 & $-$12.07  &4.7$\pm$2.2\\
1627+444   & 24$\pm$6.5& $-11.35\pm 0.13$ & 2.20$\pm$0.32 & 53.23$\pm$0.37 & 39.03 & $-$22.12 & $-$11.33 &9.6$\pm$4.0\\
1658+302   &  48$\pm$9  & $-13.05\pm 0.17$ & 0.31$\pm$0.06 & 50.89$\pm$0.20 & 36.10 & $-$23.67 & $-$13.12  &5.2$\pm$1.5\\
1723+510   &  20$\pm$3  & $-11.54\pm 0.18$ & 1.76$\pm$0.36 & 53.48$\pm$0.27 & 39.55 & $-$22.29 & $-$11.53 &16.0$\pm$7.7\\
1726+318   &  27$\pm$5  & $-12.48\pm 0.14$ & 0.60$\pm$0.10 & 52.75$\pm$0.29 & 38.14 & $-$23.76 & $-$12.67  &4.5$\pm$2.1\\
1957+405   &  13$\pm$3  & $-10.39\pm 0.15$ & 6.60$\pm$1.18 & 53.64$\pm$0.25 & 39.39 & $-$22.03 & $-$10.61  &4.7$\pm$1.6\\
2019+098   & 9.2$\pm$1.2& $-11.14\pm 0.10$ & 2.80$\pm$0.32 & 53.33$\pm$0.27 & 39.28 & $-$23.20 & $-$11.22  &5.6$\pm$2.7\\
2104+763   & 7.7$\pm$0.6& $-10.87\pm 0.14$ & 3.80$\pm$0.63 & 53.31$\pm$0.31 & 39.06 & $-$23.28 & $-$11.25  &2.9$\pm$1.9\\
2145+151   &  6.4$\pm$1.1& $-10.75\pm 0.13$ & 4.38$\pm$0.64 & 53.61$\pm$0.38 & 39.87 & $-$22.70 & $-$10.90 &7.4$\pm$3.9\\
\hline
\end{tabular*}
\end{table*}

\begin{table}[htb]
\caption{Observational and model parameters characterizing the source's
cocoon}
\begin{tabbing}
xxxxxxxxxxxxxxxxxxxxxxxxxxxxxxxxxxxxxx\=xxxxxxxxxxxx\= \kill 
Parameter                     \>Symbol    \>Dimension\\
\end{tabbing}

{\sc Observational parameters from radio maps and spectra}
\begin{tabbing}
xxxxxxxxxxxxxxxxxxxxxxxxxxxxxxxxxxxxxx\=xxxxxxxxxxxx\= \kill
Apparent (projected) linear size \> $D$      \>[kpc]\\
axial ratio                      \> $AR$     \>[dimensionless]\\
observed volume                  \> $V_{0}$  \>[kpc$^{3}$]\\
1.4-GHz luminosity               \> $P_{\rm 1.4}$\>[W\,Hz$^{-1}$sr$^{-1}$]\\
source redshift                  \> $z$      \>[dimensionless]\\
\end{tabbing}

{\sc Physical parameters derived directly from the above data}
\begin{tabbing}
xxxxxxxxxxxxxxxxxxxxxxxxxxxxxxxxxxxxxx\=xxxxxxxxxxxx\= \kill
equipartition magnetic field strength \> $B_{\rm eq}$ \> [nT]\\
equipartition energy density     \> $u_{\rm eq}$ \> [J\,m$^{-3}$]\\
equipartition emitted energy     \> $U_{\rm eq}\equiv u_{\rm eq}V_{0}$ \> [J]\\
\end{tabbing}

{\sc Physical parameters assumed in the model}
\begin{tabbing}
xxxxxxxxxxxxxxxxxxxxxxxxxxxxxxxxxxxxxx\=xxxxxxxxxxxx\= \kill
Central core radius              \> $a_{0}$  \> [kpc]\\
exponent of density profile      \> $\beta$  \> [dimensionless]\\
exponent of particle energy distribution \> $p$      \> [dimensionless]\\
ratio of jet-head to cocoon pressure \> ${\cal P}_{\rm hc}\equiv p_{\rm h}/p_{\rm c}$\>[dimensionless]\\
adiabatic indices of ambient medium,\\
 cocoon,
and magnetic field, respectively \>$\Gamma_{\rm x}$, $\Gamma_{\rm c}$, $\Gamma_{\rm B}$\>[dimensionless]\\
source axis inclination          \>$\theta$   \>[$^{\circ}$]\\
\end{tabbing}

{\sc Physical parameters fitted with the model for a given age}
\begin{tabbing}
xxxxxxxxxxxxxxxxxxxxxxxxxxxxxxxxxxxxxx\=xxxxxxxxxxxx\= \kill
Jet (constant) power             \>$Q_{0}$    \>[W]\\
central core density             \>$\rho_{0}$ \>[kg\,m$^{-3}$]\\
cocoon pressure                  \>$p_{\rm c}$\>[N\,m$^{-2}$]\\
energy density                   \>$u_{\rm c}$\>[J\,m$^{-3}$]\\
total emitted energy             \>$E_{\rm tot}\equiv u_{\rm c}V_{\rm c}$\>[J]
\end{tabbing}
\end{table}

\begin{table}[h]
\caption{The correlations between (log) $D$ and $t$ or $Q_{0}$ or $\rho_{0}$
when other parameters are kept constant}
\begin{tabular}{lclrll}
\hline
Correlation & $r_{XY}$ & $r_{XY/U}$  & $P_{XY/U}$\\
            & & $r_{XY/V}$  & $P_{XY/V}$\\
            & & $r_{XY/W}$  & $P_{XY/W}$  & $r_{XY/UVW}$  & $P_{XY/UVW}$\\
\hline
$D-t$/$Q_{0}$ &+0.804       & +0.941  & $\ll$0.001\\
$D-t$/$\rho_{0}$   &  & +0.805  & $\ll$0.001\\
$D-t$/1+$z$        &  & +0.839  & $\ll$0.001\\
$D-t$/$Q_{0}$,$\rho_{0}$,1+$z$  & & &  & +0.952  & $\ll$0.001\\
& &\\
$D-Q_{0}$/$\rho_{0}$ &$-$0.065 & +0.015  & 0.898\\
$D-Q_{0}$/$t$      &  & +0.810  & $\ll$0.001\\
$D-Q_{0}$/1+$z$    &  & +0.450  & 0.001\\
$D-Q_{0}$/$\rho_{0}$,$t$,1+$z$ & &  & & +0.866  & $\ll$0.001\\
& &\\
$D-\rho_{0}$/$Q_{0}$ &$-$0.211 & $-$0.256 & 0.031\\
$D-\rho_{0}$/$t$   &  & $-$0.080 & 0.505\\
$D-\rho_{0}$/1+$z$ &  & $-$0.125 & 0.298\\
$D-\rho_{0}$/$Q_{0}$,$t$,1+$z$  & & & & $-$0.716 & $\ll$0.001\\
\hline
\end{tabular}
\end{table}

\noindent
In view of the dynamical model applied and as a result of the above statistical
correlations, we see that the linear size of a source strongly depends on
both its age and the jet power, where the correlation with age is the strongest.
However, the size also anti-correlates with central density of the core. That
anticorrelation seems to be a weaker than the correlations with $Q_{0}$ and $t$  
and become well pronounced only when
all three remaining parameters ($Q_{0}$, $t$ and $z$) are kept constant.

Fitting a surface to the values of $D$ over the $t$--$Q_{0}$ plane, we find

\begin{equation}
D(t,Q_{0})\propto t^{1.06\pm 0.03}Q_{0}^{0.31\pm 0.02}.
\end{equation}

\noindent
The above relation well illustrates the influence of the jet power on the source
(cocoon) expansion velocity, i.e. its length at a given age. However, the source
expansion velocity ought to depend also on the external environment conditions.
Indeed, the significant partial correlation coefficient for the $D-\rho_{0}$
correlation in Table~4 confirms this effect.

A greater than one exponent of the age in Equation~(10) may suggest a
statistical acceleration of the expansion velocities with age which contradicts
with the deceleration implied by Equation~(1) for $\beta$=1.5 (shown with the
dashed line in Figure~3). However, the observed luminosities of the sample sources
demand different jet powers (cf. see below). A linear regression of
their apparent $D$ values transformed to a constant reference value of $Q_{0}$
with Equation~(10) on the age axis gives $D(t)\propto t^{1.05\pm 0.05}$. This is
somehow a surprised result and there would be several explanations of it: (i)
$\beta>2$, (ii) non-constant $Q_{0}$ during the lifetime of sources, (iii) not
representative sample of sources, or (iv) real possibility of an acceleration of
the expansion speed, at least for sources evolving in a specific environmental
conditions. We will return to this possibility in Paper~II.
 
The partial correlation coefficients calculated for the correlation between
the luminosity $P_{\rm 1.4}$ and $t$, $Q_{0}$ and 1+$z$, as well as the related
probabilities of their chance correlations are given in Table~5.

\begin{figure}[t]
\includegraphics{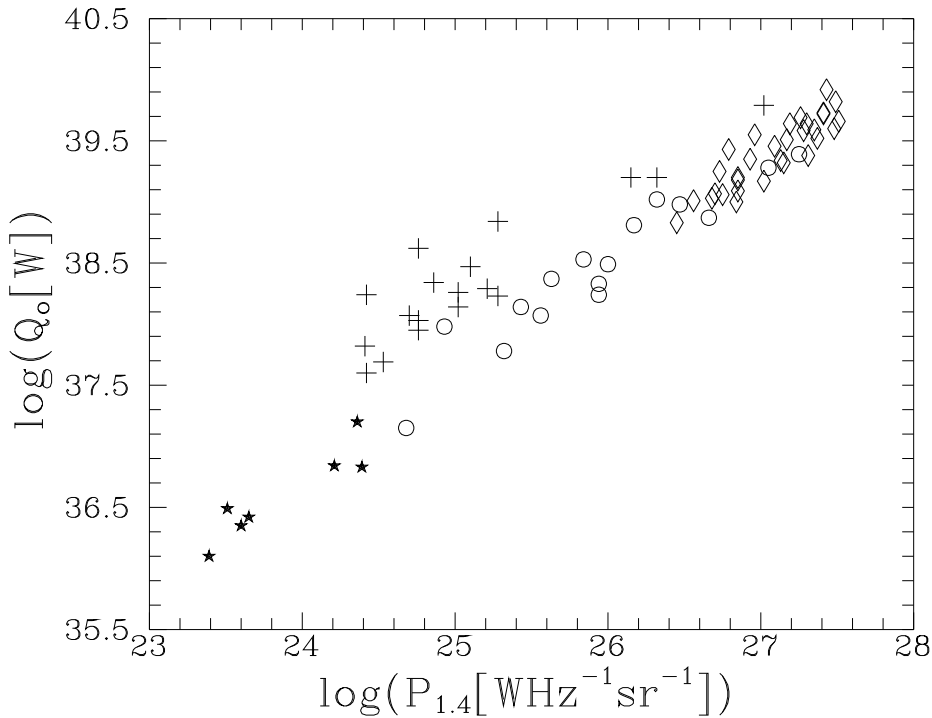}
\vspace{75mm} 
\caption{Plot of the model jet power $Q_{0}$ against the observed 1.4 GHz luminosity
of the sample sources.}
\label{f4}
\end{figure}

\begin{table}[h]
\caption{The correlations between (log) $P_{\rm 1.4}$ and $t$ or $Q_{0}$ or 1+$z$
when other parameters are kept constant}
\begin{tabular}{lclrlr}
\hline
Correlation & $r_{XY}$ & $r_{XY/U}$  & $P_{XY/U}$\\
            & & $r_{XY/V}$  & $P_{XY/V}$\\
            & &             &             & $r_{XY/UV}$  & $P_{XY/UV}$\\
\hline
$P_{\rm 1.4}-t$/$Q_{0}$ &$-$0.718  &$-$0.647  & 0.001\\
$P_{\rm 1.4}-t$/1+$z$   &  &$-$0.351  & 0.002\\
$P_{\rm 1.4}-t$/$Q_{0}$,1+$z$  & & &  &$-$0.623  & 0.005\\
& &\\
$P_{\rm 1.4}-Q_{0}$/$t$ &+0.940  & +0.926  & $\ll$0.001\\
$P_{\rm 1.4}-Q_{0}$/1+$z$ & & +0.854  & $\ll$0.001\\
$P_{\rm 1.4}-Q_{0}$/$t$,1+$z$ & & &  & +0.900  & $\ll$0.001\\
& &\\
$P_{\rm 1.4}-$(1+$z$)/$Q_{0}$ &+0.771& +0.270 & 0.023\\
$P_{\rm 1.4}-$(1+$z$)/$t$  &   & +0.515 & $<$0.001\\
$P_{\rm 1.4}-$(1+$z$)/$Q_{0}$,$t$ & & &  & $-$0.153 & 0.21\\
\hline
\end{tabular}
\end{table}

\noindent
Table~5 shows that the strongest correlation is between the source apparent
luminosity and its jet power, although the anticorrelation between the luminosity
and age is also significant. The model values of $Q_{0}$ vs. the observed 1.4-GHz
luminosities of the sample sources are plotted in Figure~4.

\begin{figure}[t]
\includegraphics{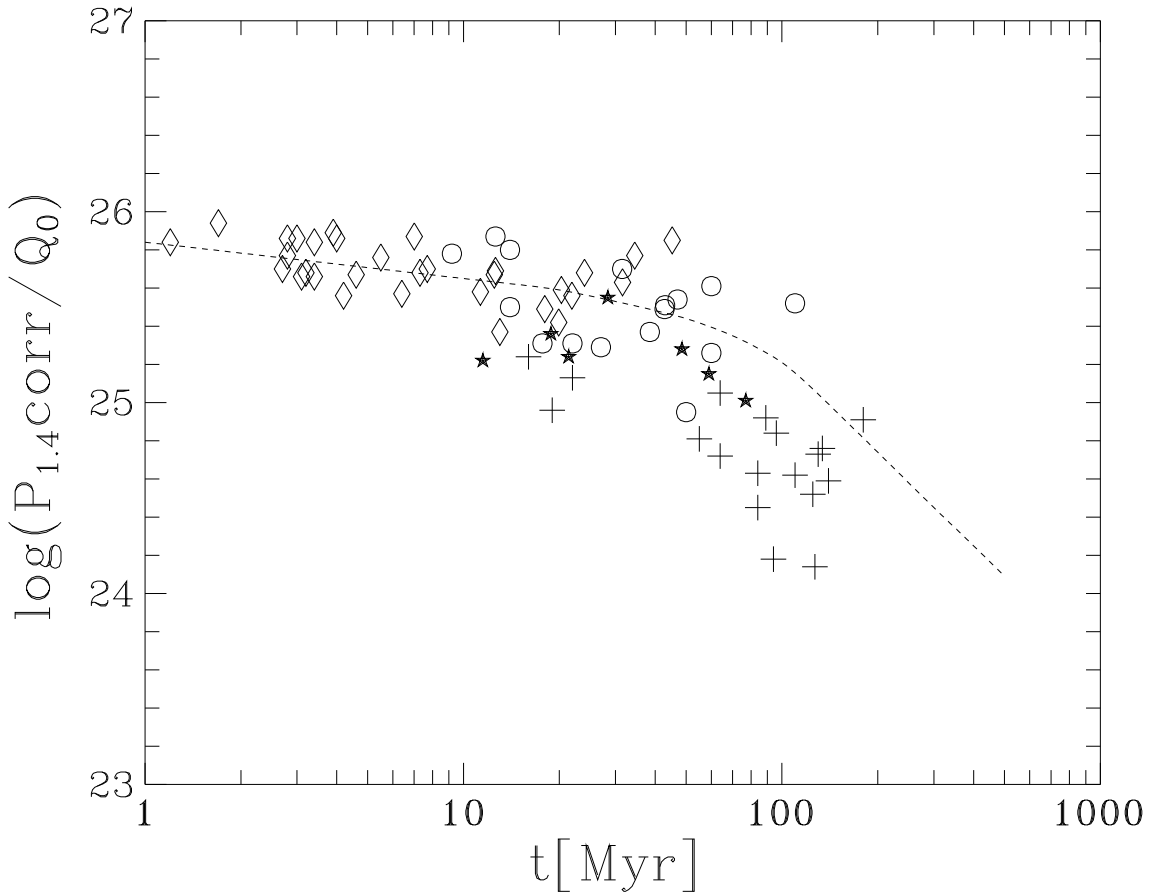}
\vspace{75mm} 
\caption{Plot of the 1.4 GHz luminosity $P_{\rm 1.4}$ transformed to a constant
jet power of 10$^{38}$\,W against the source age $t$. The symbols indicating sources
are the same as in Figure~1. The dashed curve shows the $P-t$ relation for 
$Q_{0}$=10$^{38}$\,W resulting 
from the iterative solution of the relevant KDA equations; cf. Section~3.3.}
\label{f5}
\end{figure}

The {\sl giant} sources seemingly have either a higher jet powers or lower radio
luminosities than the normal-size sample sources. According to the model applied,
the source luminosity (at a constant $Q_{0}$) decreases with time, thus Figure~4
confirms that the {\sl giants}, statistically older than normal-size sources in the
sample (cf. Table~2), are less luminous than younger comparison sources with a
similar $Q_{0}$.

Fitting a surface to the values of $P_{\rm 1.4}$ over the $t$--$Q_{0}$ plane, we find

\[P_{\rm 1.4}(t,Q_{0})\propto t^{-0.61\pm 0.08}Q_{0}^{0.99\pm 0.05}.\]

\noindent
An anticorrelation between apparent luminosity and age of matured sources has been
predicted by the KDA and BWR analytical models. The data of the sample sources
allow to verify those predictions. This anticorrelation in our sample is shown in
Figure~5 where the 1.4-GHz luminosities are transformed to the constant jet power
of 10$^{38}$\,W according to the surface fit given above. There is an evident lack
of powerful old sources concordant with predictions of the above two models. 
However, the
luminosity of {\sl giant} sources seems to decrease faster with respect to that
predicted by the KDA model and very likely is connected with a departure from the
self-similarity of the cocoon expansion in that model. Therefore our data on
{\sl giant} sources support rather the evolutionary predictions of the BWR model.
 
The first time (to our knowledge) axial ratios of {\sl giant} sources were
analysed and compared with those of smaller FRII-type sources by Subrahmanyan et
al. (1996), who found no difference between the axial ratios of eight {\sl
giants} and eight 3C sources with a median size of about 400 kpc. The authors
did not specify which 3C sources were considered, but since all {\sl giant} and
normal sources were of comparable powers and at comparable redshifts, we
assume they might be of similar ages, so the dependence of $AR$ on time could
not be visible.
 
In the BRW model  the axial ratio of an individual source
steadily increases throughout its lifetime. Moreover, that model implies a
dependence of the $AR$ on the jet power $Q_{0}$. The latter dependence was
probably reflected by an apparent correlation between $AR$ and the 178-MHz
luminosity
of 3C sources noted by Leahy and Williams (1984). Taking into account the
unavoidable anticorrelation between $Q_{0}$ and age in any sample of sources (cf.
Section~4.1), in Table~6 we
have calculated the partial correlation coefficients and the related
probabilities of chance correlations between $AR$ and $t$, $AR$ and $Q_{0}$, and
$AR$ and $\rho_{0}$ when relevant combinations of the parameters $t$, $Q_{0}$,
$\rho_{0}$, and 1+$z$ are kept constant. 

\begin{table}[htb]
\caption{The correlations between (log) $AR$ and $t$ or $Q_{0}$ or $\rho_{0}$
when the other parameters are kept constant}
\begin{tabular}{@{}lclrlr}
\hline
Correlation & $r_{XY}$ & $r_{XY/U}$  & $P_{XY/U}$\\
            &  & $r_{XY/V}$  & $P_{XY/V}$\\
            &  & $r_{XY/W}$  & $P_{XY/W}$  & $r_{XY/UVW}$  & $P_{XY/UVW}$\\
\hline
$AR-t$/$Q_{0}$ &+0.328   & +0.615  & $\ll$0.001\\
$AR-t$/$\rho_{0}$ &  & +0.445  & $<$0.001\\
$AR-t$/1+$z$      &  & +0.488  & $<$0.001\\
$AR-t$/$Q_{0}$,$\rho_{0}$,1+$z$ & &  &  & +0.513  & $<$0.001\\
& &\\
$AR-Q_{0}$/$t$ &+0.235    & +0.570 & $\ll$0.001\\
$AR-Q_{0}$/$\rho_{0}$ & & +0.121 & 0.314\\
$AR-Q_{0}$/1+$z$   & & +0.445 & $<$0.001\\
$AR-Q_{0}$/$t$,$\rho_{0}$,1+$z$  & & &  & +0.421 & $<$0.001\\
& &\\
$AR-\rho_{0}$/$t$ &+0.225       & +0.380  & 0.001\\
$AR-\rho_{0}$/$Q_{0}$    & & +0.182  & 0.129\\
$AR-\rho_{0}$/1+$z$      & & +0.301  & 0.011\\
$AR-\rho_{0}$/$t$,$Q_{0}$,1+$z$  & & &  & +0.209  & 0.085\\
\hline
\end{tabular}
\end{table}

Table~6 shows statistically significant correlations between the axial ratio
and the source's age, as well as the axial ratio and the jet power.
Fitting a surface to the values of $AR$ over the $Q_{0}$--$t$ plane (where
$Q_{0}$ is in watts and $t$ in Myr), we found

\begin{equation}
AR(t,Q_{0} )\propto t^{0.23\pm 0.03}Q_{0}^{0.12\pm 0.02}.
\end{equation}

\noindent
Indeed, our statistical data strongly support the implication of the BRW model of
the dependence of $AR$ on $Q_{0}$. A consequence of this effect for the expansion
speed of the cocoon is pointed out in the next subsection.
Using the above relation, we transform the apparent $AR$ values from 
Table~1 to a reference jet power of $10^{39}$\,W. The relation between the
transformed axial ratio and age of the sample sources with the regression line
on the time axis is shown in Figure~6.

\begin{figure}[t]
\includegraphics{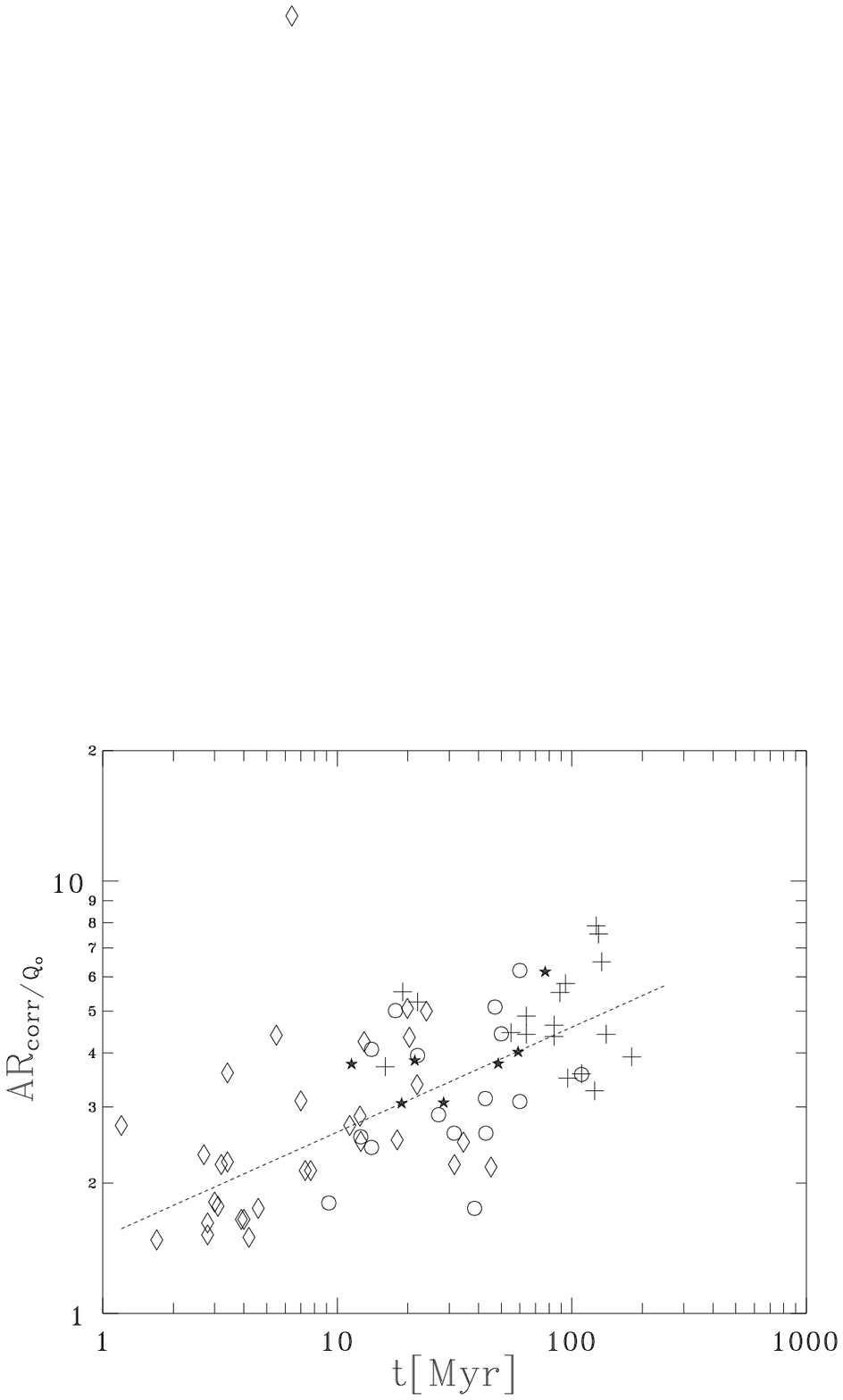}
\vspace{75mm} 
\caption{Plot of the source axial ratio transformed to a constant jet power of
10$^{39}$\,W against the age $t$. The dashed line indicates the linear regression
on the age axis.}
\label{f6}
\end{figure}

This statistical correlation between the cocoon's axial ratio and age
implies a time evolution of the ratio of the pressure in the head
of the jet to the cocoon pressure. Indeed, substitution of Equation (11) into
Equation (3) (for $\beta$=1.5 with its assumed uncertainty of $\pm$0.4) gives:

\begin{equation}
{\cal P}_{\rm hc}(t,Q_{0})\propto t^{0.38\pm 0.06}Q_{0}^{0.21\pm 0.04}.
\end{equation}

\noindent
which violates the model assumption of self-similar expansion of the cocoon. The
consequence of this will be analysed in Paper II.

\begin{figure}
\includegraphics{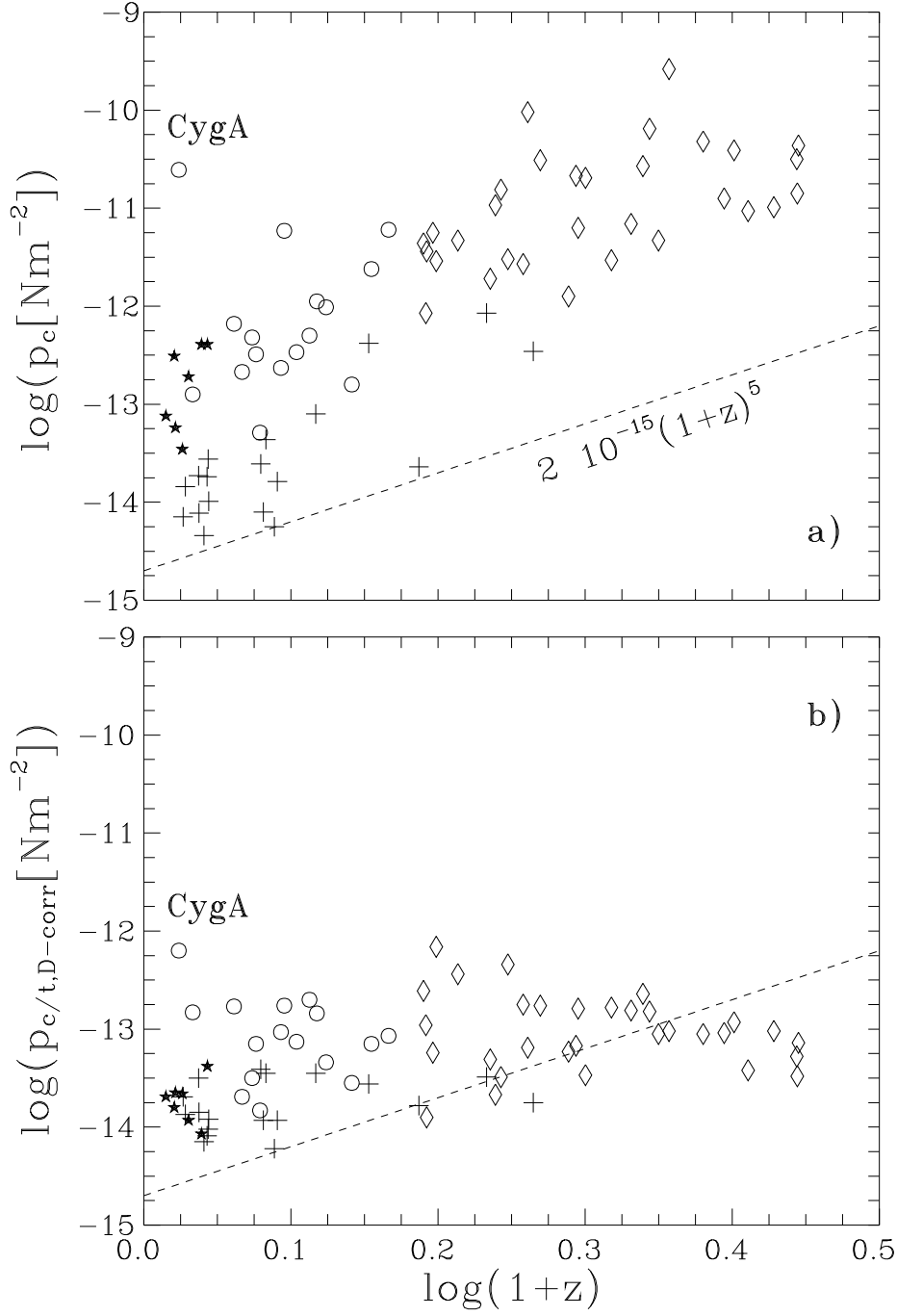}
\vspace{155mm}
\caption{{\bf a)} Cocoon pressure against size of the sample sources; {\bf b)}
the same pressure transformed to the reference size of 1 Mpc and age of 100 Myr
against redshift. The symbols indicating sources within different sets are the
same as in Figure 1 and 2. The dashed line in {\bf a} and {\bf b}
indicates the pressumed IGM presure evolution $p_{\rm IGM}\propto (1+z)^{5}$.}
\label{f7}
\end{figure}

Our statistical analysis confirms the significant anticorrelation between the
cocoon pressure and the size of sources expected from Equation~(2), but reveals
also  a significant correlation between this pressure and redshift. Besides, the
size strongly correlates with age and anticorrelates with redshift, so we have
calculated the partial correlation between all these parameters. The Pearson
partial correlation coefficients between $p_{\rm c}$, $D$, $t$, and 1+$z$ are
given in Table~7.

\begin{table}
\caption{The correlation between (log) $p_{\rm c}$ and $D$ or $t$ or 1+$z$ where
the other parameters are kept constant}
\begin{tabular}{@{}lclrlr}
\hline
Correlation & $r_{XY}$  & $r_{XY/U}$  & $P_{XY/U}$\\
              &    & $r_{XY/V}$  & $P_{XY/V}$\\
       &  & & & $r_{XY/UV}$  & $P_{XV/UV}$\\
\hline
$p_{\rm c}-D/t$ &$-$0.792  & $-$0.234    & 0.051\\
$p_{\rm c}-D$/1+$z$ & & $-$0.806  & $\ll$0.001\\
$p_{\rm c}-D/t$,1+$z$  & &  &    & $-$0.533  & $<$0.001\\
&\\
$p_{\rm c}-t/D$ &$-$0.917  & $-$0.772    & $\ll$0.001\\
$p_{\rm c}-t$/1+$z$ & & $-$0.826  & $\ll$0.001\\
$p_{\rm c}-t/D$,1+$z$  & &   &    & $-$0.467  & 0.001\\
&\\
$p_{\rm c}-$(1+$z)/D$ & +0.785 & +0.733  & $\ll$0.001\\
$p_{\rm c}-$(1+$z)/t$ & & +0.145  & 0.23\\
$p_{\rm c}-$(1+$z)/D,t$ & &  &   & +0.324 & 0.005\\
\hline
\end{tabular}
\end{table}

Table~7 shows that: (i) The correlation coefficients for the direct ($r_{XY}$) and
partial ($r_{XY/V}$, $r_{XY/UV}$) correlations between the cocoon pressure and its
size, and the cocoon pressure and the
source age are very similar. This is obvious in view of the very high size--age
correlation shown in Table~4. (ii) Both direct correlations are weakened, though
still remain very significant, if the age and redshift (in the $p_{\rm c}-D$
correlation) and the size and redshift (in the $p_{\rm c}-t$ correlation) are kept
constant. Formally, the strongest partial correlation is found between the cocoon
pressure and size (in fact: the cocoon's volume). (iii) the strong direct correlation
between the cocoon pressure and redshift is seriously weakened if the source size
and age are kept constant. This correlation, shown in Figure~7a, is very important
for studies of physical conditions in the intergalactic medium (IGM).
The dashed line indicates the expected electron
pressure in an adiabatically expanding Universe in the form $p_{\rm IGM}=
p^{0}_{\rm IGM}(1+z)^{5}$ with $p^{0}_{\rm IGM}=2\cdot10^{-15}$ N\,m$^{-2}$ (cf.
Subrahmanyan and Saripalli 1993).

Fitting a
surface to the values of $p_{\rm c}$ over the $D-t$ plane we find

\[p_{\rm c}(D,t)\propto D^{-0.30\pm 0.19}t^{-1.78\pm 0.17}.\]

\noindent
Therefore, one can transform the cocoon's pressure values to a reference size and
age. The plot of cocoon pressures transformed to $D$=1 Mpc and $t$=100 Myr versus
1+$z$ is shown in Figure~7b. This Figure illustrate how $p_{\rm c}$ of the 
high-redshift sources would decrease if they had evolved into the above size and
age, and emphasize that the strong direct $p_{\rm c}-$(1+$z$) correlation in
Figure~7a is, in fact, due to the much stronger correlations $p_{\rm c}-D$ and
$p_{\rm c}-t$ (cf. Table~7). However, this is purely statistical result and the
question about conditions under which distant radio sources would reach very large
size and old age remains open.  The aspect of a
cosmological evolution of the IGM is discussed in Section~5.4.

\section{Discussion of the Results}

\subsection[section]{Influence of Fixed Model Parameters on the Model Predictions}

The basic physical parameters of the radio sources derived in this paper with the
aim of the KDA model, i.e. jet
power $Q_{0}$, central density $\rho_{0}$, and cocoon pressure $p_{\rm c}$, are
in principle dependent on the assumed central core radius, the exponent in the
external gas distribution, the adiabatic indices of electrons and magnetic field,
as well as on the orientation of the jet axis towards the observer. In application
of the KDA model we have assumed the same values of these parameters for all sample
sources. This assumption can only be valid in our statistical analysis of the
evolutionary trends in the whole FRII-type population but not for individual
sample sources. In particular, we have adopted $a_{0}=10$ kpc and $\beta=1.5$,
respectively.   Taking  a lower
core radius, e.g. $a_{0}=2$ kpc, and keeping $\beta=1.5$ will result in increase
of the model density of the core $\rho_{0}$ by roughly one order of
magnitude, while other model parameters will not be changed. Conversely, a
lower density gradient, e.g. $\beta=1.1$, will lower $\rho_{0}$ approximately
$1.5\sim 6$ times. In this case however, the jet power will be increased by a
few percent and accordingly the pressure and energy density in the cocoon will be
changed. Solving the equations in Section~3 for a few sets of the model free
parameters ($a_{0}$, $\beta$, $\theta$) we find that the presented and discussed
correlations between the observational and model parameters are not changed; all
the statistical trends are preserved although the values of the model parameters
(especially $\rho_{0}$) are changed quantitatively. 

A relativistic equation of state for the magnetic field does not
change our results significantly, unless the KDA `Case~1' ($\Gamma_{\rm c}=\Gamma_
{\rm B}=4/3$) is adopted. However, this case, i.e. when both the cocoon and the
magnetic field energy  have a relativistic equation of state, is unlikely for
our sample sources. Therefore, we argue that the results discussed below are not
significantly biased by selection of a particular set of the model parameters.

\subsection[section]{Cause of Extremal Linear Size}

In view of the KDA and BRW analytical models of dynamical evolution of FRII-type
radio sources, many such sources can evolve into a stage characterized by a
linear size exceeding 1 Mpc. Access to this stage depends on a number of the
model parameters: jet power $Q_{0}$, its Lorentz factor $\gamma_{\rm jet}$, the
adiabatic indices of the cocoon material and magnetic field, $\Gamma_{\rm c}$
and $\Gamma_{\rm B}$, respectively, as well as the core radius $a_{0}$, the external
gas density $\rho_{0}$, and the exponent of its distribution $\beta$. For a given
set of these parameters, the model allows us to determine whether an evolving
source will reach  the size of 1 Mpc, and if so, at what age. From this
point of view, {\sl giant} sources should be the oldest ones.

In this paper (Section 4.3) we have confronted the KDA model predictions with 
the observational data on {\sl giant}-size and normal-size FRII sources. Our 
statistical analysis strongly suggests that there is not a single evolutionary 
scheme governing the size development. An old age or a low external density 
alone is insufficient to assure extremely large linear extent of a source, 
both are necessary together with a suitable power driven from AGN by the 
highly relativistic jets. The Pearson partial correlation analysis indicates
that the dependence of linear size on each of these three parameters is
statistically significant. Ordering these correlations by decreasing partial
correlation coefficients, we find that the size is dependent on age, then on
$Q_{0}$; next on $\rho_{0}$.

About 83 \% of {\sl giants}
in our sample possess a projected linear size over 1 Mpc owing to statistically
old age, low or moderate density of the external medium, and high enough power
of their jets. The remaining 17 \% (3 sources) are high luminosity 
sources at redshifts $z>0.4\sim 0.5$ and ages from 15 to 25 Myr which are
typical for normal-size sources. Two of them are quasars. The jet power of these
sources is high enough to compensate for a higher ram pressure in a denser
surrounding environment and higher energy losses during the cocoon expansion.
The jet power of {\sl giants} is not extreme, so several FRII-type sources
having that $Q_{0}$ can potentially achieve very large size after a suitably
long time.
According to our results (cf. Figure~1a), these potential {\sl giants} should
have $Q_{0}>10^{37.5}$ W and be situated in an environment with $\rho_{0}<10^{-23}$
kg\,m$^{-3}$.

The above scenario is not caused by a selection effect. We show that there are
low-luminosity sources which lie in the parts of $\log Q_{0}-\log\rho_{0}$
plane (Figure~1a) completely avoided
by {\sl giant} sources. They diverge from {\sl giants} and even normal-size
powerful sources by having low jet power $Q_{0}<10^{37.5}$ W and total energy
$E_{\rm tot}<10^{52.5}$ J. Thus we suggest that they never reach the giant size.
It is worth noting that some of them already have an age comparable to that of
typical {\sl giants}, in accordance with the model expectations.

\subsection[section]{Jet Power and Energy Budget}

The ratio of total energy supplied by the twin jets during the lifetime of a
source $(2Q_{0}t)$ and its energy stored in the cocoon, derived from the data
under the assumption of energy equipartition, $U_{\rm eq}=u_{\rm eq}V_{\rm o}$,
allow another test of the dynamical model predictions. The ratio of
$2Q_{0}t/U_{\rm eq}$ for the sample sources (given in column 9 of Table~2) is
plotted in Figure~8 vs. the cocoon axial ratio $AR$. The
uncertainties in both values are marked by error bars on some data points.
The solid curve indicates the model prediction from Equation (8), while the dashed
curve shows the best fit to the data. The observed trend fully corresponds to
the model prediction. However, the derived values of $2Q_{0}t/U_{\rm eq}$, i.e.
the reciprocal of the efficiency factor by which the kinematic energy of the jets
is converted into radiation, is much higher than $\sim$2, a value usually
assumed in a number of papers.

\begin{figure}
\includegraphics{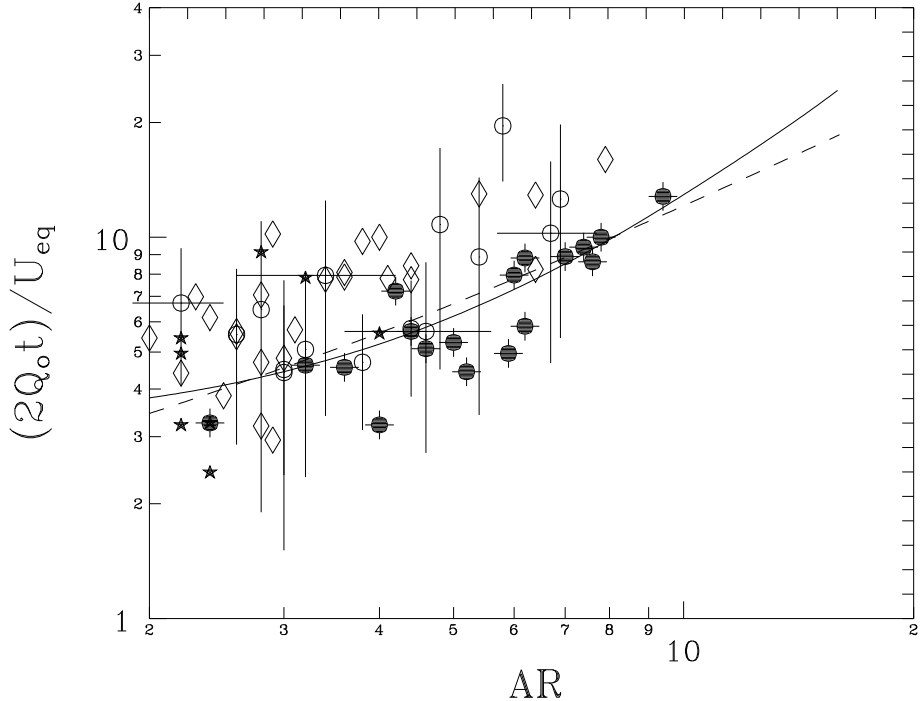}
\vspace{75mm}
\caption[]{Ratio of the total energy supplied by twin jets and energy stored
in the cocoon against its axial ratio $AR$. The large uncertainties in both
parameters are marked by error bars for a few sources only for clarity. Here the
giant sources are marked with filled circles. The solid curve indicates the model
prediction from Equation (8); the dashed curve shows the best
fit to the weighted data.}
\label{f8}
\end{figure}

In a number of studies of {\sl giant} radio sources (e.g. Parma et al. 1996;
Schoenmakers et al. 1998) the authors followed the approach of Rawlings and Saunders
(1991) and assumed a fraction of the total jet energy wasted for adiabatic expansion
of the cocoon to be about 0.5 and used it to estimate the jet power $Q_{0}$ for
sources with known age (almost always from spectral ageing analysis). In the
KDA model the energy stored in the source (cocoon) is

\[E_{\rm tot}\approx\int\{Q_{0}\,dt-(p_{\rm c}\,d[V_{\rm c}(t)]+
p_{\rm h}\,d[V_{\rm h}(t)])\} \]

\noindent
where $p_{\rm c}\,dV_{\rm c}+p_{\rm h}\,dV_{\rm h}$ is the work done to expand
the cocoon, and $p_{\rm h}$ and $V_{\rm h}$ are the hotspot pressure and volume,
respectively. If $V_{\rm h}$ is neglected, the expansion
work will be $\approx 0.5\,Q_{0}t$; if not, it is dependent on the pressure ratio
${\cal P}_{\rm hc}$  as shown in Section~3.2.

The values of $2Q_{0}t/U_{\rm eq}$ in our sample (Figure~8; Table~2) vary from 
about 2 to more than 10 due to the high jet power required for the adiabatic expansion
of the cocoon. The $Q_{0}$ values derived here from the KDA model can be 
compared with the relevant values estimated by Wan, Daly and Guerra (2000) [WDG]
for 22 3C sources included also in our sample (21 with $z$$>$0.5 + Cyg\,A). 
Recalculating their values for $H_{0}$=50 km\,s$^{-1}$Mpc$^{-1}$ we found the WDG
estimates approximately 2.5 times lower than the KDA values. Exactly the similar
ratio (1.7$\div$5.5 depending on the value of ${\cal P}_{\rm hc}$)  characterize
the jet powers of 9 3C sources, common with our sample sources, determined by
Rawlings and Saunders (1991). The explanation of this ratio is straightforward.
The values of $Q_{0}$ estimated in those papers were based on the ram pressure
considerations in the overpressured source model A of Scheuer (1974) and its
further modifications (e.g. Begelman and Cioffi 1989; Loken et al. 1992; Nath 1995).
They all are self-similar models of the Carvaldo and O'Dea type I which describe
the source dynamics only. If the source energetics and especially the energy losses
are properly taken into account (the type III models; e.g. KDA, BRW), the
significantly higher values of $Q_{0}$ are implied.
 
The data in our sample show a
dependence of the energy ratio $2Q_{0}t/U_{\rm eq}$ on the cocoon axial ratio
$AR$, and imply an increase of the fraction of jet energy spent on the adiabatic
expansion of the cocoon volume in time. However, the data also suggest that for
a constant $AR$ (i.e. a given geometry of the cocoon), {\sl giants} tend to have
a smaller ratio of $2Q_{0}t/U_{\rm eq}$ than normal-size sources which means less
energy of the jets converted into adiabatic expansion of the cocoon. This
may indicate a lower pressure of the external medium surrounding the {\sl giant}
sources than that around smaller ones.

\subsection[section]{External Pressure of the Surrounding Medium and its Evolution}

A non-relativistic uniform intergalactic medium (IGM) in thermal equilibrium
filling an adiabatically expanding Universe should have an electron pressure
evolving with redshift $p_{\rm IGM}=p^{0}_{\rm IGM}(1+z)^{5}$. The advancing
hotspots of FRII-type radio sources are probably confined by ram pressure of the
IGM. {\sl Giant} sources, with their lobes extended far outside typical galaxy
halo, have the lowest values of $p_{\rm c}$ and may be useful for determining the
upper limit of $p_{\rm IGM}$. Using a small sample of {\sl giant} sources,
Subrahmanyan and Saripalli (1993) limited its local value to
 $p^{0}_{\rm IGM}\approx(0.5\div 2)\cdot10^{-15}$ N\,m$^{-2}$. A
further study was undertaken by Cotter (1998) who, using a larger sample of 7C
giants with sources out to redshift of $\sim$0.9, confirmed a strong dependence
of the lowest $p_{\rm c}$ on redshift in agreement with a $(1+z)^{5}$ relation.
This observational result has been critically discussed by Schoenmakers et al.
(2000), who have considered possible selection effects in Cotter's analysis 
(including the Malmquist bias),
and concluded that there was not evidence in their own sample for a cosmological
evolution of $p_{\rm IGM}$. However, they also state that this hypothesis
cannot be rejected until some low-pressure high-redshift sources are found.

In all the above analyses the age of sources was not considered. The very
significant correlation between $p_{\rm c}$ and $D$ or $t$, shown in Section~4.3,
strongly suggests that the intrinsic dependences of size on age as well as of
age on redshift and not the
Malmquist bias is mainly responsible for the apparent correlation between
$p_{\rm c}$ and 1+$z$. Nevertheless, we agree with Schoenmakers et al.'s
conclusion that until {\sl giant} sources with internal pressures in their lobes
$p_{\rm c}<2\cdot10^{-15}$ N\,m$^{-2}$ at redshifts of at least $0.6\div 0.8$ are
not discovered, the IGM pressure evolution in the form
$p_{\rm IGM}\propto (1+z)^{5}$ cannot be rejected.

Most {\sl giants} in our sample, except four high-redshift ones, reveal the
lowest pressure in their cocoons. Sharing Subrahmanyan and Saripalli's arguments,
we can expect those cocoon pressures are
indicative of an upper limit to the present-day external pressure of the IGM,
$p_{\rm IGM}^{0}$. Taking into account the lowest values of $p_{\rm c}$ (cf.
Figure~7a), we found $p_{\rm IGM}^{0}<2\cdot10^{-15}$ N\,m$^{-2}$ in
accordance with their value.
It is worth emphasizing that the above results are obtained from the
analytical model assuming energy equipartition in the initial ratio of the energy
densities of the magnetic field and the particles (cf. Section~3.1). This may
not be the case in every part of the source (cocoon). Hardcastle and Worrall (2000)
estimated gas pressures in the X-ray-emitting medium around normal-size 3CRR
FRII radio galaxies and quasars, and found that, with few exceptions, the
minimum pressures in their lobes, determined under equipartition conditions,
were well below the environment pressures measured from old ROSAT observations.
Therefore, they have argued that there must be an additional contribution to the
internal pressure in lobes of those sources likely including pressure from protons,
magnetic fields exceeding their minimum-energy values, or non-uniform filling
factors.
Nevertheless, the diffuse lobes of {\sl giants}, extending farther from a host
galaxy than the typical radius of high-density X-ray-emitting gas,
may be in equilibrium with an ambient medium the emissivity of which is not
directly detectable.

\section{Conclusions}

In this paper we confront the analytical KDA model predictions with the 
observational data of `giant' and normal size FRII-type radio sources.
From our analysis we can conclude as follows:

(1) {\sl Giant}  sources do not form a separate class of radio sources, and
do not reach their extremal sizes exclusively due to some exceptional physical
conditions of the external medium. The size is dependent, in order of decreasing
partial correlation coefficients, on age; then on the jet power $Q_{0}$; next
on the central core density $\rho_{0}$.

(2) {\sl Giants} possess the lowest equipartition magnetic field strength and
energy density of their cocoons making their detection difficult in
synchrotron emission. However, their accumulated total energy is the highest
among all sources and exceeds $3\cdot10^{52}$ W.

(3) Our data confirm the conclusion drawn by Blundell et al. (1999) that
throughout the lifetime of an individual source its axial ratio can steadily
increase, thus its expansion cannot be self-similar all the time. A
self-similar expansion seems to be feasible if the power supplied by the jets
is a few orders of magnitude above the minimum-energy value. In other cases
the expansion can only initially be self-similar; a departure from
self-similarity for large and old sources is justified by observations of
{\sl giant} sources.

(4) The apparent increase of the lowest cocoon pressures (observed in the
largest sources) with redshift is mainly caused by the intrinsic dependence of
their age on redshift and dominates over a bias by possible selection effects.
However, a
cosmological  evolution of the IGM cannot be rejected until {\sl giant} sources
with internal pressures in their lobes less than $2\cdot10^{-15}$ N\,m$^{2}$ at
high redshifts are {\sl not} discovered.

\section{Acknowledgements}

We thank Dr. C. R. Kaiser for explanations of the integration procedures used in
the KDA paper.

\newpage

\begin{center}
{Observational data and physical parameters of the sample sources}
\end{center}

Table~1 gives the observational data of the sample sources.
Their radio luminosity, size and volume are calculated with $H_{0}$=50 
km\,s$^{-1}$Mpc$^{-1}$ and $q_{0}$=0.5. The entries are:

{\sl Column 1:} IAU-format name of source

{\sl Column 2:} Other name

{\sl Column 3:} Redshift

{\sl Column 4:} Logarithm of 1.4 GHz luminosity

{\sl Column 5:} Projected linear size

{\sl Column 6:} Source axial ratio

{\sl Column 7:} Logarithm of source volume

{\sl Column 8 and 9:} References to the radio maps and spectral ageing analysis,
respectively 

\vspace{3mm}
Table~2 contains the adopted age and physical parameters of the sample
sources derived either from the observational data (columns 3, 4, and 5) and
from the analytical KDA model (columns 6, 7, 8, and 9). The entries are:

{\sl Column 2:} Age of source

{\sl Column 3:} Logarithm of the equipartition energy density

{\sl Column 4:} Equipartition magnetic field strength 

{\sl Column 5:} Logarithm of the source total energy

{\sl Column 6:} Logarithm of the jet power

{\sl Column 7:} Logarithm of the initial density of the
external medium at the core radius $a_{0}$

{\sl Column 8:} Logarithm of the cocoon pressure

{\sl Column 9:} Ratio between the total energy of twin jets and the source energy.


\begin{thebibliography}{}

\bibitem{alex}
{Alexander, P., and Leahy J.P.} {1987,} {\it MNRAS,} {225,} {1}
\bibitem{barthel}
{Barthel, P.D.} {1989,} {\it ApJ,} {336,} {606}
\bibitem{becker}
{Becker, R.H., White, R.L., and Helfand, D.J.} {1995,} {\it ApJ,} {450,} {559}
\bibitem{begelman}
{Begelman, M.C., and Cioffi, D.F.} {1989,} {\it ApJ,} {345,} {L21}
\bibitem{blundell}
{Blundell, K.M., Rawlings, S., and Willott, C.J.} {1999,} {\it AJ,} {117,} {766}
\bibitem{canizares}
{Canizares, C.R., Fabbiano, G., and Trincheiri, G.} {1987,} {\it ApJ,} {312,} {503}
\bibitem{carilli}
{Carilli, C.L., Perley, R.A., Dreher, J.W., and Leahy, J.P.} {1991,} {\it ApJ,} {383,} {554}
\bibitem{carvalho}
{Carvalho, J.C., and O'Dea, C.P.} {2002,} {\it ApJS,} {141,} {337}
\bibitem{condon}
{Condon, J.J., Cotton, W.D., Greisen, E.W., Yin, Q.F., Perley, R.A., 
Taylor,~G.B., and Broderick,~J.J.} {1998,} {\it AJ,} {115,} {1693}
\bibitem{cotter}
{Cotter, G., 1998} {``Observational cosmology with the new sky surveys''}{~}
{Eds.: M.N. Bremer et al. (Kluwer Acad. Publ.), p.233}
\bibitem{daly}
{Daly, R.A.} {1995,} {\it ApJ,} {454,} {580}
\bibitem{ruiter}
{de Ruiter, H.R., Parma, P., Fanti, C., and Fanti, R.} {1986,} {\it A\&AS,} {65,} {111}
\bibitem{fanaroff}
{Fanaroff, B.L., and Riley, J.M.} {1974,} {\it MNRAS,} {167,} {31P}
\bibitem{fanti}
{Fanti, C., Fanti, R., de Ruiter, H.R., and Parma, P.} {1986,} {\it A\&A,} {65,} {145} 
\bibitem{guerra}
{Guerra, E.J., Daly, R.A., and Wan, L.} {2000,} {\it ApJ,} {544,} {659}
\bibitem{hardcastle}
{Hardcastle, M.J., and Worrall, D.M.} {2000,} {\it MNRAS,} {319,} {562}
\bibitem{hine}
{Hine, R.G.} {1979,} {\it MNRAS,} {189,} {527}
\bibitem{ishwara}
{Ishwara-Chandra, C.H., and Saikia, D.J.} {1999,} {\it MNRAS,} {309,} {100}
\bibitem{jaffe}
{Jaffe, W.J., and Perola, G.C.} {1973,} {\it A\&A,} {26,} {423}
\bibitem{kaiser1}
{Kaiser, C.R.} {2000,} {\it A\&A,} {362,} {447}
\bibitem{kaiser2}
{Kaiser, C.R., and Alexander, P. (KA)} {1997,} {\it MNRAS,} {286,} {215}
\bibitem{kaiser3}
{Kaiser, C.R., and Alexander, P.} {1999,} {\it MNRAS,} {302,} {515}
\bibitem{kaiser4}
{Kaiser, C.R., Dennett-Thorpe, J., and Alexander, P. (KDA)} {1997,} {\it MNRAS,} {292,} {723}
\bibitem{king}
{King, I.R.} {1972,} {\it ApJ,} {174,} {L123}
\bibitem{klein}
{Klein, U., Mack, K.-H., Gregorini, L., and Parma, P.} {1995,} {\it A\&A,} {303,} {427}
\bibitem{lara}
{Lara, L., Mack, K.-H., Lacy, M., Klein, U., Cotton, W.D., Feretti, L.,
Giovannini, G., and Murgia, M.} {2000,} {\it A\&A,} {356,} {63}
\bibitem{loken}
{Loken, C., Burns, J.O., Clarke, D.A., and Norman, M.L.} {1992,} {\it ApJ,} {392,} {54}
\bibitem{leahy1}
{Leahy, J.P., Muxlow, T.W.B., and Stephens, P.W.} {1989,} {\it MNRAS,} {239,} {401}
\bibitem{leahy2}
{Leahy, J.P., and Williams, A.G.} {1984,} {\it MNRAS,} {210,} {929}
\bibitem{liu}
{Liu, R., Pooley, G., and Riley, J.M.} {1992,} {\it MNRAS,} {257,} {545}
\bibitem{mach1}
{Machalski, J., and Condon, J.J.} {1983,} {\it AJ,} {88,} {1591}
\bibitem{mach2}
{Machalski, J., and Condon, J.J.} {1985,} {\it AJ,} {90,} {5}
\bibitem{mach3}
{Machalski, J., and Jamrozy, M.} {2000,} {\it A\&A,} {363,} {L17}
\bibitem{mach4}
{Machalski, J., Jamrozy, M., and Zola, S.} {2001,} {\it A\&A,} {371,} {445}
\bibitem{mack1}
{Mack, K.-H., Klein, U., O'Dea, C.P., and Willis, A.G.} {1997,} {\it A\&AS,} {123,} {463}
\bibitem{mack2}
{Mack, K.-H., Klein, U., O'Dea, C.P. Willis, A.G., and Saripalli, L.} {1998,}
{\it A\&A,} {329,} {431}
\bibitem{miley}
{Miley, G.K.} {1980,} {\it ARA\&A,} {18,} {165}
\bibitem{myers}
{Myers, S.T., and Spangler, S.R.} {1985,} {\it ApJ,} {291,} {52}
\bibitem{nath}
{Nath, B.B.} {1995,} {\it MNRAS,} {274,} {208} 
\bibitem{parma1}
{Parma, P., de Ruiter, H.R., Fanti, C., and Fanti, R.} {1986,} {\it A\&AS,} {64,} {135}
\bibitem{parma2}
{Parma, P. de Ruiter, H.R., Mack, K.-H., van Breugel, W., Dey, A., Fanti,~R.,
and Klein, U.} {1996,} {\it A\&A,} {311,} {49}
\bibitem{parma3}
{Parma, P., Murgia, M., Morganti, R., Capetti, A., de Ruiter, H.R.,
and Fanti,~R.} {1999,} {\it A\&A,} {344,} {7}
\bibitem{rawlings}
{Rawlings, S., and Saunders, R.} {1991,} {\it Nature,} {349,} {138}
\bibitem{saripalli}
{Saripalli, L., Subrahmanyan, R., and Hundstead, R.W.} {1994,} {\it MNRAS,} {269,} {37}
\bibitem{scheuer}
{Scheuer, P.A.G.} {1974,} {\it MNRAS,} {166,} {513}
\bibitem{schoen1}
{Schoenmakers, A.P., de Bruyn, A.G., R\"{o}ttgering, H.J.A., and van der Laan,~H.}
{2001,} {\it A\&A,} {374,} {861}
\bibitem{schoen2}
{Schoenmakers, A.P., Mack, K.-H., Lara, L., R\"{o}ttgering, H.J.A.,
de Bruyn,~A.G., van der Laan,~H., and Giovannini, G.} {1998,} {\it A\&A,} {336,} {455}
\bibitem{schoen3}
{Schoenmakers, A.P., Mack, K.-H., de Bruyn,~A.G., R\"{o}ttgering, H.J.A., 
Klein,~U., and van der Laan,~H.} {2000,} {\it A\&AS,} {146,} {293}
\bibitem{spangler}
{Spangler, S.R., and Pogge, J.J.} {1984,} {\it AJ,} {89,} {342}
\bibitem{subrah1}
{Subrahmanyan, R., and Saripalli, L.} {1993,} {\it MNRAS,} {260,} {908}
\bibitem{subrah2}
{Subrahmanyan, R., Saripalli, L., and Hundstead, R.W.} {1996,} {\it MNRAS,} {279,} {257}
\bibitem{urry}
{Urry, C.M., and Padovani, P.} {1995,} {\it PASP,} {107,} {803}
\bibitem{wan}
{Wan, L., Daly, R.A., and Guerra, E.J.} {2000,} {\it ApJ,} {544,} {671}
\bibitem{wellman}
{Wellman, G.F., Daly, R.A., and Wan, L.} {1997,} {\it ApJ,} {480,} {96}

\end{thebibliography}
\end{document}